\PassOptionsToPackage{table}{xcolor}
\documentclass[12pt]{article}

\usepackage{amsfonts,amsmath,amssymb}
\usepackage{hyperref}
\usepackage{cite}
\usepackage{epsfig}
\usepackage{paralist}
\usepackage{fancyhdr}
\usepackage{tikz}
\usepackage{tkz-euclide}
\usetikzlibrary{decorations.pathmorphing,patterns,calc,snakes,arrows}

\usepackage{graphicx}
\usepackage{xcolor}
\numberwithin{equation}{section}
\usepackage[vcentermath]{youngtab}
\usepackage{etex}
\usepackage{braket}
\usepackage{float}
\usepackage{placeins}

\def\spa#1{\phantom{\fbox{\rule[-#1cm]{0cm}{0cm}}}}

\def\be{\begin{equation}}
\def\ee{\end{equation}}
\def\bea{\begin{eqnarray}}
\def\eea{\end{eqnarray}}

\renewcommand{\thefootnote}{\fnsymbol{footnote}}

\makeatletter
\g@addto@macro\bfseries{\boldmath}
\makeatother

\def\Tb{{\bar{T}}}

\def\dim{\mathop{\mathrm{dim}}\nolimits}

\begin{document}

\hfuzz=100pt
\title{{\Large \bf{A Third-Quantized Description of Spacetime Wormholes in AdS/CFT}}}
\date{}
\author{Shinji Hirano$^{a, b}$\footnote{
	e-mail:
	\href{mailto:shinji.hirano@gmail.com}{shinji.hirano@gmail.com}}}
\date{}

\maketitle

\thispagestyle{fancy}
\rhead{YITP-26-113}
\cfoot{}
\renewcommand{\headrulewidth}{0.0pt}

\vspace*{-1cm}
\begin{center}
$^{a}${{\it School of Science, Huzhou Normal University}}
\\ {{\it Huzhou 313000, Zhejiang, China}}
  \spa{0.5} \\
$^b${{\it Center for Gravitational Physics and Quantum Information (CGPQI)}}
\\ {{\it  Yukawa Institute for Theoretical Physics, Kyoto University}}
\\ {{\it Kitashirakawa-Oiwakecho, Sakyo-ku, Kyoto 606-8502, Japan}}
\spa{0.5}  

\end{center}

\begin{abstract}
We propose an extension of AdS/CFT in which quantum gravitational
wavefunctions of connected bulk geometries are assembled into a Fock
space of universes. Splitting and joining interactions in a
third-quantized Hamiltonian then provide a quantum description of
changes in bulk connectivity and topology. Reduced phase-space
quantization supplies the physical one-universe Hilbert space and a
radial Schr\"odinger evolution with respect to which the
topology-changing interactions are ordered. We develop the construction
concretely in AdS$_3$ gravity with a torus boundary, where these
ingredients can be described explicitly. The light spectrum is treated
as fixed, non-normalizable background data, whereas the normalizable
states above the black-hole threshold are quantized as dynamical modes.
A conventional boundary theory is selected by choosing a coherent state
whose universe-field expectation value, together with the fixed
background contribution, yields consistent CFT data at the asymptotic
boundary. Generic one-universe wavefunctions need not admit such an
interpretation. 
This raises the possibility that CFT-realizable wavefunctions are
nongeneric in the gravitational state space and are not closed under
arbitrary superpositions of gravitational states.
The known two-torus spacetime-wormhole amplitude, which lies at the heart of the
factorization problem, provides concrete input for the
topology-changing interactions of the third-quantized theory. A
complementary cutting-and-gluing description yields an effective
ensemble interpretation when the resulting gravitational sectors admit
consistent boundary-theory interpretations, without assuming a
fundamental ensemble of boundary theories. We also discuss replica
wormholes and baby-universe processes within the same framework.
\end{abstract}

\renewcommand{\thefootnote}{\arabic{footnote}}
\setcounter{footnote}{0}

\newpage

\tableofcontents


\section{Introduction}
\label{sec:Introduction}

AdS/CFT is conventionally formulated for a fixed boundary theory and
bulk geometries with prescribed asymptotic boundaries
\cite{Maldacena:1997re, Witten:1998qj, Gubser:1998bc}. 
Quantum gravity, however, naturally admits
processes in which connected spatial components split or join and the
topology of the bulk changes. A complete gravitational description must therefore account not only
for quantum evolution within a sector of fixed spatial topology, but
also for transitions between sectors containing different numbers and
connectivities of spatial components.

Spacetime wormholes provide a particularly sharp manifestation of this
problem. Smooth Euclidean AdS wormholes with several asymptotic
boundaries produce connected bulk contributions that appear
incompatible with factorization in a fixed product of boundary
theories, as emphasized by Maldacena and Maoz
\cite{Maldacena:2004rf}.\footnote{A related but distinct factorization
problem arises in the canonical quantization of two-sided JT gravity:
the constrained gravitational Hilbert space does not naturally
factorize into independent left and right boundary factors
\cite{Harlow:2018tqv}. This Hilbert-space factorization problem should
be distinguished from the nonfactorizing multi-boundary amplitudes
generated by spacetime wormholes.}

A tractable setting for investigating this problem emerged from the
convergence of two developments. The first was Jackiw--Teitelboim (JT)
gravity \cite{Jackiw:1984je,Teitelboim:1983ux}, a two-dimensional
dilaton-gravity theory in which gravitational path integrals with
nontrivial topology can be studied explicitly. The second was the
Sachdev--Ye--Kitaev (SYK) model. The disordered Sachdev--Ye model
\cite{Sachdev:1992fk} and its Majorana-fermion formulation introduced
by Kitaev \cite{KitaevTalks} provided a solvable many-body system with
black-hole-like thermodynamics and quantum-chaotic spectral
statistics. The identification of their common low-energy Schwarzian
dynamics connected the SYK model to the boundary dynamics of
nearly-AdS$_2$ gravity
\cite{Maldacena:2016hyu,Jensen:2016pah,Maldacena:2016upp}.

More broadly, the relation between black holes and quantum chaos had
been sharpened through the analysis of scrambling and the butterfly
effect \cite{Shenker:2013pqa}. The subsequent identification of
random-matrix-like spectral correlations in black-hole systems,
particularly through the late-time spectral form factor
\cite{Cotler:2016fpe}, provided an important precursor to the
gravitational interpretation of the ramp and plateau in terms of
wormhole geometries.

Together, these developments made it possible to ask whether the ramp
in the ensemble-averaged SYK spectral form factor has a direct
gravitational origin. Saad, Shenker, and Stanford identified the
corresponding connected saddle as the double-cone wormhole
\cite{Saad:2018bqo}. Its structure and normalization were subsequently
examined in greater detail in Ref.~\cite{Chen:2023hra}. The double cone
thereby provides a gravitational realization of correlations normally
associated with random-matrix level statistics. The subsequent
identification of the perturbative topological expansion of JT gravity
with that of a double-scaled matrix integral made the relation between
gravitational wormholes and random-matrix observables precise
\cite{Saad:2019lba}.

Nevertheless, the precise status of the putative gravitational
ensemble remains unclear. The matrix-integral description does not by
itself determine whether ensemble averaging is fundamental, represents
a form of coarse-graining, or provides an effective organization of
observables in an underlying single theory. 
Kudler-Flam and Witten proposed instead that the relevant smooth gravitational data may be extracted by Mellin averaging over a canonically specified large-\(N\) family of CFTs. This averages across distinct finite-\(N\) theories, but does not require a separately postulated random ensemble at fixed \(N\)~\cite{Kudler-Flam:2025cki,Kudler-Flam:2026nzz}. 
Liu developed this idea into a large-$N$ filtering proposal that removes erratic
microscopic dependence while retaining the smooth contributions and
correlations associated with semiclassical wormhole
amplitudes~\cite{Liu:2025ikq,Liu:2026fnd}.\footnote{A related role for large-$N$ averaging arises in the
closed-universe setting: Gesteau's no-go theorem excludes a
semiclassical baby universe entangled with the two exterior regions
under its low-energy and ordinary large-$N$ convergence assumptions,
while convergence only after averaging can evade this
conclusion~\cite{Gesteau:ClosedUniverses}.}
Independently, Perlmutter used the Fyodorov--Hiary--Keating conjecture
to bound erratic fluctuations of high-energy CFT spectra across
energy windows at fixed $N$, finding a residual of order
$S_0 e^{-S_0}$ relative to the semiclassical density and thereby a
precision limit for the semiclassical gravitational path integral; he
further proposes a translation of this energy-window erraticness into
the large-$N$ erraticness addressed by
Refs.~\cite{Kudler-Flam:2025cki,Kudler-Flam:2026nzz,Liu:2025ikq,Liu:2026fnd,Perlmutter:BlackHoles}.

The present work addresses a further level of the problem. Rather than
treating one-universe wavefunctions only as boundary observables, we
promote them to quantum fields. The one-universe Hilbert space becomes
the one-particle space of a universe Fock space, while splitting and
joining interactions generate transitions between sectors with
different universe number. Connected multi-boundary amplitudes may
then arise dynamically as correlations generated by the
topology-changing Hamiltonian, without requiring a fundamental
ensemble of boundary theories.

This possibility has several important precedents. Coleman introduced
creation and annihilation operators for baby universes and the
associated $\alpha$-state sectors to describe wormhole-induced effects
on parent-universe physics
\cite{Coleman:1988cy,Coleman:1988tj}. This quantization of the
baby-universe sector was extended by Giddings and Strominger to a more
general third-quantized framework in which complete universes undergo
splitting and joining processes \cite{Giddings:1988wv}. The relation
between baby-universe sectors, boundary factorization, and
ensemble-like observables was subsequently sharpened by Marolf and
Maxfield \cite{Marolf:2020xie}.

More recently, the universe field theory of JT gravity was formulated
as a Kodaira--Spencer theory on the matrix-model spectral curve
\cite{Post:2022dfi}. Its perturbative expansion reproduces the
gravitational sum over topologies through interactions describing the
splitting and joining of entire JT universes. Thus, even in a setting
where the matrix-integral description makes an ensemble interpretation
particularly natural, the same topology-changing amplitudes admit a
third-quantized organization.

In this work, we develop a third-quantized extension of AdS/CFT based
on reduced phase-space quantization. We begin with a physical
gravitational one-universe Hilbert space and promote its normalizable
wavefunctions to creation and annihilation modes in a universe Fock
space. Because the gravitational constraints have already been solved,
the free theory exhibits first-order Schr\"odinger evolution in a
radial time. Topology-changing cobordisms are represented by
interaction vertices connecting sectors with different universe
number, and their consistency is constrained by requiring amplitudes
to be independent of their decomposition into elementary splitting and
joining processes.

The resulting construction may be viewed as a reduced-phase-space
analogue of the temporal-gauge noncritical string field theory
developed by Ishibashi and Kawai and related work
\cite{Ishibashi:1993nq,Fukuma:1993np,Ikehara:1994vx}. In that
two-dimensional setting, spatial loops are treated as the fundamental
quanta, and a Hamiltonian with splitting and joining vertices
generates worldsheets of varying topology. Here, complete connected
gravitational hypersurfaces replace the spatial loops, while radial
Hamiltonian evolution generates spacetime histories of varying
connectivity. The construction also differs from conventional third
quantization based directly on the Wheeler--DeWitt equation: the
one-universe wavefunctions used here are physical states obtained
after solving the constraints, and their topology-changing dynamics is
organized by a first-order evolution operator.

The construction becomes explicit in AdS$_3$ gravity with torus
boundary topology. In this sector, the normalizable physical
one-universe Hilbert space admits an orthogonal modular spectral
decomposition, and its radial evolution is known. The vacuum and light
spectrum are treated as fixed non-normalizable background data,
whereas the normalizable modes above the black-hole threshold are
promoted to dynamical universe modes. The known two-torus
spacetime-wormhole amplitude \cite{Cotler:2020ugk} then provides
concrete input for the topology-changing dynamics. 
We show that its diagonal modular spectral representation is
compatible with a splitting interaction satisfying the
third-quantized sewing conditions. This constrains, but does not
uniquely determine, the topology-changing dynamics.

The ordinary fixed-topology description is embedded into the universe
Fock space by choosing an appropriate coherent state. Its universe
field expectation value, together with the fixed non-normalizable
contribution, reproduces the complete one-universe wavefunction. This
coherent state plays a role analogous to a Coleman $\alpha$-state and
to the definite-theory sectors appearing in the Marolf--Maxfield
construction \cite{Marolf:2020xie}, although the analogy is not exact.
It is an eigenstate of the universe annihilation operators and evolves
under the topology-changing Hamiltonian into a correlated, generally
non-Gaussian multi-universe state. Unlike the boundary-generated
construction of Ref.~\cite{Marolf:2020xie}, the one-universe states
used here are physical states before the Fock space is formed.

The framework gives a common description of several apparently
different gravitational phenomena. Spacetime-wormhole correlations
are obtained as connected universe-field correlators in the evolved
coherent state. Replica wormholes arise as particular $n$-point
amplitudes selected by cyclic sewing of cut-open boundary data.
Cutting a spacetime wormhole in a complementary direction exposes an
entangled two-universe state and leads, under additional conditions,
to an effective ensemble representation. The statistical
interpretation is therefore a property of particular amplitudes and
states within the third-quantized theory, rather than the definition
of the underlying gravitational theory. The same splitting
interactions also provide a natural setting for discussing
baby-universe production.

The construction raises a broader question about the scope of
holography. Standard discussions of AdS/CFT fix a boundary theory and
study the bulk states represented in its Hilbert space. Here, by
contrast, the physical space of one-universe wavefunctions is
constructed intrinsically on the gravitational side. One may then ask
the inverse question: whether every admissible complete gravitational
wavefunction is realized by a consistent boundary theory. The torus
example makes this issue concrete because its physical modes and their
radial evolution are known explicitly, whereas a particular CFT
selects only a distinguished combination through its boundary
partition function. We return to the relation between gravitational
wavefunctions, CFT realizability, and the selection of gravitational
initial conditions in Subsection~\ref{sec:CFT}.

The remainder of this paper is organized as follows. 
In Section~\ref{sec:Wormhole3rdQ}, we formulate the universe Fock
space and its topology-changing Hamiltonian, introduce the sewing and
Schwinger--Dyson constraints on multi-universe amplitudes, and describe
spacetime-wormhole and replica-wormhole correlations. We also discuss
their effective ensemble interpretation and implications for AdS/CFT.
In Section~\ref{sec:AdS3}, we implement the construction in the
AdS$_3$ torus sector and show that the leading two-torus wormhole
amplitude is compatible with a splitting interaction satisfying the
third-quantized sewing condition.
Section~\ref{sec:higherD} discusses possible extensions
to higher-dimensional gravitational wavefunctions. Finally, in
Section~\ref{sec:discussion}, we consider broader implications and open problems,
including the interpretation of baby-universe processes. The reduced
phase-space quantization and York-time evolution underlying the
construction are reviewed in Appendix~\ref{sec:CQ_York}.


\section{Third Quantization, Topology Change, and Spacetime Wormholes}
\label{sec:Wormhole3rdQ}

Canonical quantization ordinarily begins by fixing the topology of a
codimension-one hypersurface $\Sigma$ and quantizing the gravitational
degrees of freedom defined on it. The resulting Hilbert space
$\mathcal H_{\Sigma}$ describes quantum states of a universe with that
fixed topology. Evolution generated within $\mathcal H_{\Sigma}$ may
change the geometry of $\Sigma$, but it cannot describe a process in
which $\Sigma$ changes topology or splits into several connected
components. Its Hamiltonian description therefore requires an
enlargement of the canonical state space.

Topology-changing geometries may instead be included covariantly
through gravitational path integrals over cobordisms. A conventional
path integral, however, is defined for prescribed boundary data and
therefore fixes both the topology and the number of its boundary
components. It may sum over bulk topologies compatible with those
data, but it does not by itself sum over amplitudes with different
numbers of exposed boundaries. An $m\to n$ process, in which an
initial hypersurface with $m$ connected components evolves into one
with $n$ connected components, must be defined by a path integral over
the corresponding cobordisms. Different choices of $(m,n)$ therefore
give a family of multi-boundary amplitudes, which third quantization
assembles into a single dynamical description.

A crucial ingredient in passing from these multi-boundary amplitudes
to a third-quantized Hamiltonian is the existence of a physical
evolution parameter. A covariant cobordism amplitude by itself does
not distinguish a splitting process from its reverse. The reduced
phase-space formulation adopted here supplies such a parameter: after
the gravitational constraints are solved, the physical wavefunction
undergoes first-order Schr\"odinger evolution in a radial time $T$.
This construction is reviewed in Appendix~\ref{sec:CQ_York}. The
ordering in $T$ distinguishes incoming from outgoing hypersurfaces,
orders topology-changing events, and allows cobordism amplitudes to be
interpreted as matrix elements of an evolution operator. In the
universe Fock space, topology-changing loci are represented by
interaction vertices connecting different sectors.

To incorporate such processes, we first assemble the Hilbert spaces
associated with different connected topologies into a one-universe
Hilbert space,
\begin{equation}
    \mathcal H_{\mathrm U}
    =
    \bigoplus_{\Sigma\ {\rm connected}}
    \mathcal H_{\Sigma}.
\end{equation}
States containing arbitrary numbers of connected universes then
belong to the bosonic Fock space
\begin{equation}
    \mathcal F_{\mathrm U}
    =
    \bigoplus_{N=0}^{\infty}
    \operatorname{Sym}^{N}\mathcal H_{\mathrm U}.
\end{equation}

In general, the physical one-universe wavefunction admits a
decomposition
\begin{equation}
    \Psi_{\mathrm{phys}}(q,T)
    =
    \Psi_{\mathrm{bg}}(q,T)
    +
    \Psi_{\mathrm{norm}}(q,T).
    \label{eq:wavefunction-background-split}
\end{equation}
Here $\Psi_{\mathrm{bg}}$ denotes a fixed non-normalizable contribution
and should not be confused with a background spacetime metric. It does
not belong to the physical one-universe Hilbert space. Only
$\Psi_{\mathrm{norm}}$ is promoted to a universe field operator and
acquires creation and annihilation modes; $\Psi_{\mathrm{bg}}$ remains
fixed under the third-quantized dynamics.

The distinction between normalizable and non-normalizable solutions is
fixed once the physical inner product and asymptotic boundary
conditions have been specified. The choice of
$\Psi_{\mathrm{bg}}$ as a particular representative is nevertheless
not always unique: one may shift
\begin{equation}
    \Psi_{\mathrm{bg}}
    \longrightarrow
    \Psi_{\mathrm{bg}}+\chi,
    \qquad
    \Psi_{\mathrm{norm}}
    \longrightarrow
    \Psi_{\mathrm{norm}}-\chi,
    \qquad
    \chi\in\mathcal H_{\mathrm U},
    \label{eq:background-representative-shift}
\end{equation}
without changing $\Psi_{\mathrm{phys}}$. Thus, the intrinsic object is
the affine space
$\mathfrak A_{\mathrm U}=\Psi_{\mathrm{bg}}+\mathcal H_{\mathrm U}$,
whereas the choice of its origin requires an additional convention.
In the third-quantized description, a change of representative amounts
to a coherent displacement of the normalizable universe field and
does not change physical amplitudes when the state and interaction
Hamiltonian are transformed consistently.\footnote{In the
AdS$_3$ torus example discussed below, we fix this freedom by
associating $\Psi_{\mathrm{bg}}$ with the prescribed vacuum and light
data and placing the remaining contribution in the normalizable
modular spectral sector. Only the latter spans the one-universe
Hilbert space and acquires creation and annihilation operators. This
separation accords with the proposal of Schlenker and Witten that
subthreshold observables show no effects of ensemble averaging,
whereas apparent ensemble behavior is confined to the black-hole
sector \cite{Schlenker:2022dyo}.}

Let $\{\psi_A(q,T)\}$ be a complete orthonormal basis of {\it normalizable}
one-universe wavefunctions, where $q$ denotes the reduced gravitational
degrees of freedom, $T$ is the chosen evolution parameter, and $A$
collectively labels the topology and physical quantum numbers of a
connected universe. We take the modes to satisfy the reduced
one-universe Schr\"odinger equation,\footnote{In the torus example discussed below, the York-time-dependent
prefactor in the reduced Hamiltonian can be absorbed into a
redefinition of York time, yielding an intrinsic WdW time $T$. Under
Wick rotation, the factors of $i$ acquired by the original York time
and the unscaled reduced Hamiltonian cancel, so that the Schr\"odinger
equation in $T$ takes the same form in the Lorentzian and Euclidean
descriptions. We assume that an analogous choice of intrinsic time is
available more generally, although this should be regarded as a
working assumption rather than a universal result. If residual
$T$ dependence remains, it can be incorporated through a
time-ordered evolution operator.}
\begin{equation}
    i\partial_T\psi_A(q,T)
    =
    \widehat h_{\mathrm{red}}(T)\psi_A(q,T),
    \label{eq:one-universe-schrodinger}
\end{equation}
where we use the shorthand
$\widehat h_{\mathrm{red}}(T)
\equiv h_{\mathrm{red}}
(q,-i\partial/\partial q;T)$
for the reduced Hamiltonian.
The corresponding interaction-picture universe field has the mode
expansion
\begin{equation}
    \widehat\Psi_I(q,T)
    =
    \sum_A\psi_A(q,T)a_A,
    \qquad
    \widehat\Psi_I^\dagger(q,T)
    =
    \sum_A\psi_A^*(q,T)a_A^\dagger.
    \label{eq:universe-field-expansion}
\end{equation}
The oscillators satisfy
\begin{equation}
    [a_A,a_B^\dagger]
    =
    \delta_{AB},
    \qquad
    [a_A,a_B]
    =
    [a_A^\dagger,a_B^\dagger]
    =
    0.
\end{equation}
Here and below, the sum over $A$ includes integration over any
continuous part of the one-universe spectrum:
\begin{equation}
    \sum_A
    \equiv
    \sum_{A\in\mathrm{discrete}}
    +
    \int_{\mathrm{continuous}}d\mu(A).
\end{equation}
Correspondingly, $\delta_{AB}$ denotes either a Kronecker or a Dirac
delta function, as appropriate. The operator $a_A^\dagger$ creates a
connected universe in the physical state $\psi_A$, while $a_A$
annihilates it.

The fixed non-normalizable contribution is restored only when
reconstructing the full physical wavefunction. In a chosen
third-quantized state $\vert\Phi\rangle$, one has schematically
\begin{equation}
    \Psi_{\mathrm{phys}}(q,T)
    =
    \Psi_{\mathrm{non\text{-}norm}}(q,T)
    +
    \langle\Phi|
    \widehat\Psi_I(q,T)
    |\Phi\rangle.
    \label{eq:full-wavefunction-expectation}
\end{equation}

In the Schr\"odinger picture, the complete third-quantized Hamiltonian
consists of a quadratic kinetic term and topology-changing
interactions,
\begin{equation}
    \widehat H_{\mathrm{3Q}}^{S}(T)
    =
    \widehat H_0^{S}(T)
    +
    \widehat H_{\mathrm{int}}^{S}(T).
    \label{eq:full-third-quantized-Hamiltonian}
\end{equation}
The kinetic term is obtained by promoting the reduced one-universe
Hamiltonian to the universe Fock space:
\begin{equation}
    \widehat H_0^{S}(T)
    =
    \int d\mu(q)\,d\mu(q')\,
    \widehat\Psi_S^\dagger(q)\,
    h_{\mathrm{red}}(q,q';T)\,
    \widehat\Psi_S(q'),
    \label{eq:third-quantized-kinetic-term}
\end{equation}
where
$h_{\mathrm{red}}(q,q';T)
\equiv
\langle q|\widehat h_{\mathrm{red}}(T)|q'\rangle$
is the matrix element of the reduced one-universe Hamiltonian in the
$q$ representation.

We henceforth work in the interaction picture defined by the
one-universe modes in
Eq.~\eqref{eq:universe-field-expansion}. Since these modes solve the
free reduced Schr\"odinger equation, the kinetic evolution has already
been absorbed into their $T$ dependence. The interaction-picture
Fock-space state therefore evolves according to
\begin{equation}
    i\partial_T\vert\Phi(T)\rangle_I
    =
    \widehat H_{\mathrm{int}}^I(T)
    \vert\Phi(T)\rangle_I.
    \label{eq:interaction-picture-evolution}
\end{equation}
The interaction Hamiltonian generates processes that change the
topology or the number of connected universes. In direct analogy with
many-body second quantization, its simplest splitting and joining
contribution is
\begin{align}
    \widehat H_{\mathrm{int}}^I(T)
    ={}&
    \frac{1}{2}
    \int
    d\mu(q_1)\,d\mu(q_2)\,d\mu(q_3)\,
    \mathcal V_{1\to2}(q_1,q_2;q_3;T)
    \nonumber\\
    &\times
    \widehat\Psi_I^\dagger(q_1,T)
    \widehat\Psi_I^\dagger(q_2,T)
    \widehat\Psi_I(q_3,T)
    +\mathrm{h.c.}
    +\cdots ,
    \label{eq:universe-splitting-interaction}
\end{align}
where $q_3$ labels the reduced configuration of the incoming universe,
$q_1$ and $q_2$ label those of the outgoing universes, and
$\mathcal V_{1\to2}$ is the topology-changing interaction kernel. The
three variables may belong to reduced configuration spaces associated
with different topologies, and the corresponding measures are
understood accordingly. The factor of $1/2$ accounts for the exchange
symmetry of the two outgoing universes.

Substituting Eq.~\eqref{eq:universe-field-expansion} gives
\begin{equation}
    \widehat H_{\mathrm{int}}^I(T)
    =
    \sum_{A,B,C}
    \left[
        g_{AB}{}^C(T)\,
        a_A^\dagger a_B^\dagger a_C
        +
        g^{AB}{}_{C}(T)\,
        a_C^\dagger a_Aa_B
    \right]
    +\cdots ,
    \label{eq:splitting-joining-Hamiltonian}
\end{equation}
where
\begin{align}
    g_{AB}{}^C(T)
    ={}&
    \frac{1}{2}
    \int
    d\mu(q_1)\,d\mu(q_2)\,d\mu(q_3)\,
    \mathcal V_{1\to2}(q_1,q_2;q_3;T)
    \nonumber\\
    &\times
    \psi_A^*(q_1,T)
    \psi_B^*(q_2,T)
    \psi_C(q_3,T).
    \label{eq:splitting-coupling}
\end{align}
Thus, $g_{AB}{}^C$ is the interaction kernel projected onto the
one-universe mode basis and determines the amplitude for a universe
in the state $C$ to split into universes in the states $A$ and $B$.
Hermiticity requires
\begin{equation}
    g^{AB}{}_{C}
    =
    \left(g_{AB}{}^C\right)^*,
\end{equation}
up to conventions for the one-universe inner product.

The ellipsis in
Eq.~\eqref{eq:splitting-joining-Hamiltonian} includes arbitrary
$m\to n$ vertices containing $m$ annihilation operators and $n$
creation operators:
\begin{equation}
    \widehat H_{m\to n}^I(T)
    =
    \frac{1}{m!\,n!}
    \sum_{\substack{
        A_1,\ldots,A_n\\
        B_1,\ldots,B_m
    }}
    g_{A_1\cdots A_n}{}^{B_1\cdots B_m}(T)\,
    a_{A_1}^\dagger\cdots a_{A_n}^\dagger
    a_{B_1}\cdots a_{B_m}.
    \label{eq:general-m-to-n-vertex}
\end{equation}
The coefficients are obtained by projecting the corresponding
$m\to n$ sewing kernels onto the one-universe mode basis. The complete
interaction Hamiltonian may be written schematically as
\begin{equation}
    \widehat H_{\mathrm{int}}^I(T)
    =
    \sum_{m,n\geq1}
    \widehat H_{m\to n}^I(T),
    \qquad
    \left(
        \widehat H_{m\to n}^I
    \right)^\dagger
    =
    \widehat H_{n\to m}^I.
    \label{eq:full-interaction-Hamiltonian}
\end{equation}
Thus the full kinetic dynamics is retained in
$\widehat H_{\mathrm{3Q}}^{S}$, while in the interaction picture it is
carried by the time-dependent one-universe modes and only the
topology-changing interactions act on the Fock-space state.

In this canonical sense, topology change naturally leads to third
quantization. The third-quantized Hamiltonian contains a one-universe
kinetic term, which generates evolution within each fixed sector, and
interaction vertices that connect sectors with different topologies
or different numbers of connected components.
Spacetime-wormhole contributions are then described by transition
amplitudes generated by these interactions. They are distinct from
Einstein--Rosen bridges. In the next subsection, we make this
distinction precise and show how cutting a spacetime-wormhole geometry
reveals the multi-universe state prepared by the corresponding
half-geometry.


\subsection{Einstein--Rosen Bridges and Spacetime Wormholes}
\label{sec:ER}

An Einstein--Rosen (ER) bridge is a codimension-one structure: it is
a connected spatial hypersurface $\Sigma_{\mathrm{ER}}$ embedded in a
spacetime $M$,
\begin{equation}
    \Sigma_{\mathrm{ER}}\subset M,
    \qquad
    \dim\Sigma_{\mathrm{ER}}=\dim M-1.
\end{equation}
It may connect two asymptotic regions while remaining a single
connected spatial geometry. The spatial slice of the two-sided
eternal black hole, and hence the geometric state associated with the
thermofield double (TFD), is the canonical example. An ER bridge is
therefore naturally associated with a state in the one-universe
Hilbert space.

By contrast, a spacetime wormhole\footnote{Here and throughout, a
spacetime wormhole refers to a connected codimension-zero bulk
geometry or cobordism with disconnected boundary components. It need
not be an on-shell classical solution or an isolated semiclassical
saddle. The on-shell Euclidean AdS wormholes of Maldacena and Maoz
have higher-genus boundaries, and no analogous smooth saddle exists
for torus boundaries~\cite{Maldacena:2004rf}. Of particular relevance
to this paper, the $T^2\times I$ wormhole, which will be one of our
principal examples below, is instead described by dominant off-shell
configurations, or constrained
instantons~\cite{Cotler:2020ugk}. The JT double trumpet provides
another familiar off-shell
example~\cite{Saad:2018bqo,Saad:2019lba}. Nevertheless, both give the
leading connected contributions responsible for the spectral ramp.}
is a codimension-zero geometry: it is a connected bulk spacetime
$M_{\mathrm{WH}}$ contributing to a gravitational amplitude. In the
setting relevant to the factorization problem, it has disconnected
boundary data,
\begin{equation}
    \partial M_{\mathrm{WH}}
    =
    B_1\sqcup B_2,
    \qquad
    M_{\mathrm{WH}}\ \text{connected},
\end{equation}
where $B_i$ denotes a boundary component of the bulk spacetime. Such
a geometry contributes to the connected part
$Z_{\mathrm{WH}}[B_1,B_2]$ of the two-boundary gravitational
amplitude $Z_{\mathrm{grav}}[B_1\sqcup B_2]$, rather than to the
factorized contribution $Z[B_1]Z[B_2]$.

The distinction may be summarized as follows:
\begin{center}
\renewcommand{\arraystretch}{1.25}
\begin{tabular}{|c|c|c|}
    \hline
    & Einstein--Rosen bridge & Spacetime wormhole \\
    \hline
    Geometric object
        & $\Sigma_{\mathrm{ER}}$
        & $M_{\mathrm{WH}}$ \\
    \hline
    Codimension
        & one
        & zero \\
    \hline
    Connectivity
        & spatial
        & spacetime \\
    \hline
    Natural quantum object
        & state
        & transition amplitude \\
    \hline
    Third-quantized role
        & one-universe sector
        & multi-universe interaction \\
    \hline
\end{tabular}
\end{center}

A useful way to compare these objects is to ask what state is prepared
by one half of a reflection-symmetric Euclidean geometry. Suppose that
a Euclidean spacetime $M$ is divided into two reflection-related
halves along a codimension-one hypersurface $\Sigma$,
\begin{equation}
    M
    =
    M_{1/2}\cup_{\Sigma}\overline{M}_{1/2},
\end{equation}
where $\overline{M}_{1/2}$ denotes the reflected, oppositely oriented
copy of $M_{1/2}$. The half-geometry prepares a wavefunctional on
$\Sigma$, and gluing the two halves reconstructs the original
amplitude:
\begin{equation}
    Z[M]
    =
    \langle\Psi_{M_{1/2}}
    \vert\Psi_{M_{1/2}}\rangle
    =
    \int\mathcal D\gamma_{\Sigma}\,
    \left|
        \Psi_{M_{1/2}}[\gamma_{\Sigma}]
    \right|^2,
\end{equation}
where $\gamma_{\Sigma}$ collectively denotes the induced gravitational
data on the gluing surface. The important distinction is therefore
not the abstract operation of taking a norm, which is common to both
cases, but the number and topology of the connected components of
$\Sigma$.

\begin{figure}[!h]
\vspace{.2cm}
\centering \includegraphics[height=2.5in]{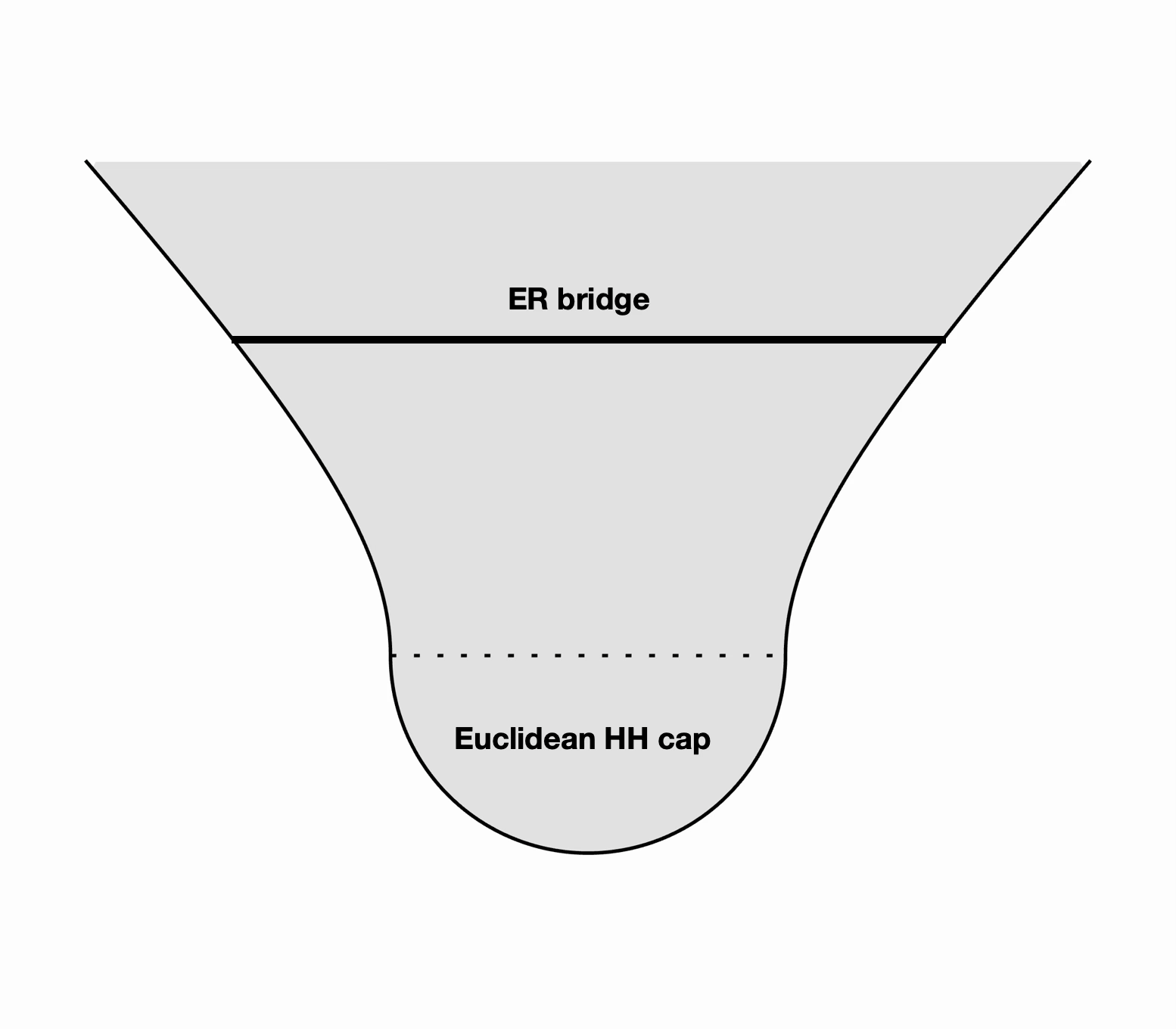}
\caption{Hartle--Hawking preparation of a single ER bridge in BTZ,
with the $S^1_\phi$ direction suppressed. The dotted line denotes the
Euclidean--Lorentzian continuation surface. The Euclidean cap prepares
the time-symmetric initial geometry containing the bifurcation surface,
which subsequently evolves in Lorentzian time. The horizontal line
represents a connected ER bridge at a later Lorentzian time, joining
the two asymptotic boundaries.}
\label{fig:ER_BTZ}
\end{figure}

The familiar example is Euclidean BTZ, whose topology is that of a
solid torus,
\begin{equation}
    M_{\mathrm{BTZ}}^{(\mathrm E)}
    \simeq
    D^2_{(r,\tau)}\times S^1_{\phi},
    \qquad
    \partial M_{\mathrm{BTZ}}^{(\mathrm E)}=T^2.
\end{equation}
The reflection-symmetric diameter of $D^2_{(r,\tau)}$ consists of the
two radial segments at
\begin{equation}
    \tau=0,
    \qquad
    \tau=\frac{\beta}{2}.
\end{equation}
Because the Euclidean-time circle contracts at the tip of the disk,
these segments meet there and form a single connected interval
$I_{\mathrm{diam}}$. The cutting surface is consequently one
connected cylinder,
\begin{equation}
    \Sigma_{\mathrm{BTZ}}
    =
    I_{\mathrm{diam}}\times S^1_{\phi}
    \simeq
    I\times S^1.
\end{equation}
Analytic continuation away from the two reflection-symmetric segments,
\begin{equation}
    \label{eq:BTZ-analytic-continuation}
    \tau=it_R,
    \qquad
    \tau=\frac{\beta}{2}+it_L,
\end{equation}
produces the right and left exterior regions of the two-sided
Lorentzian BTZ black hole. Since the two segments meet at the tip of the Euclidean disk, their
Lorentzian continuations are joined at the bifurcation surface and
together form a single connected ER bridge, as illustrated
schematically in Figure~\ref{fig:ER_BTZ}.

Each half-solid-torus therefore prepares a state on one connected ER
geometry. This is the ordinary TFD state,
\begin{equation}
    \label{eq:TFD-state}
    \vert\mathrm{TFD}(\beta)\rangle
    =
    \sum_n e^{-\beta E_n/2}
    \vert n\rangle_L\otimes\vert n\rangle_R
    \in
    \mathcal H_{\mathrm{ER}}
    \simeq
    \mathcal H_L\otimes\mathcal H_R.
\end{equation}
The reflected half prepares its conjugate, and gluing the two
half-solid-tori gives
\begin{equation}
    \label{eq:TFD-gluing}
    Z_{\mathrm{BTZ}}
    =
    \langle\mathrm{TFD}(\beta)
    \vert\mathrm{TFD}(\beta)\rangle
    =
    \sum_n e^{-\beta E_n}.
\end{equation}
Thus, the Euclidean BTZ amplitude is the norm of a one-ER state.

\begin{figure}[!h]
\vspace{.2cm}
\centering \includegraphics[height=2.5in]{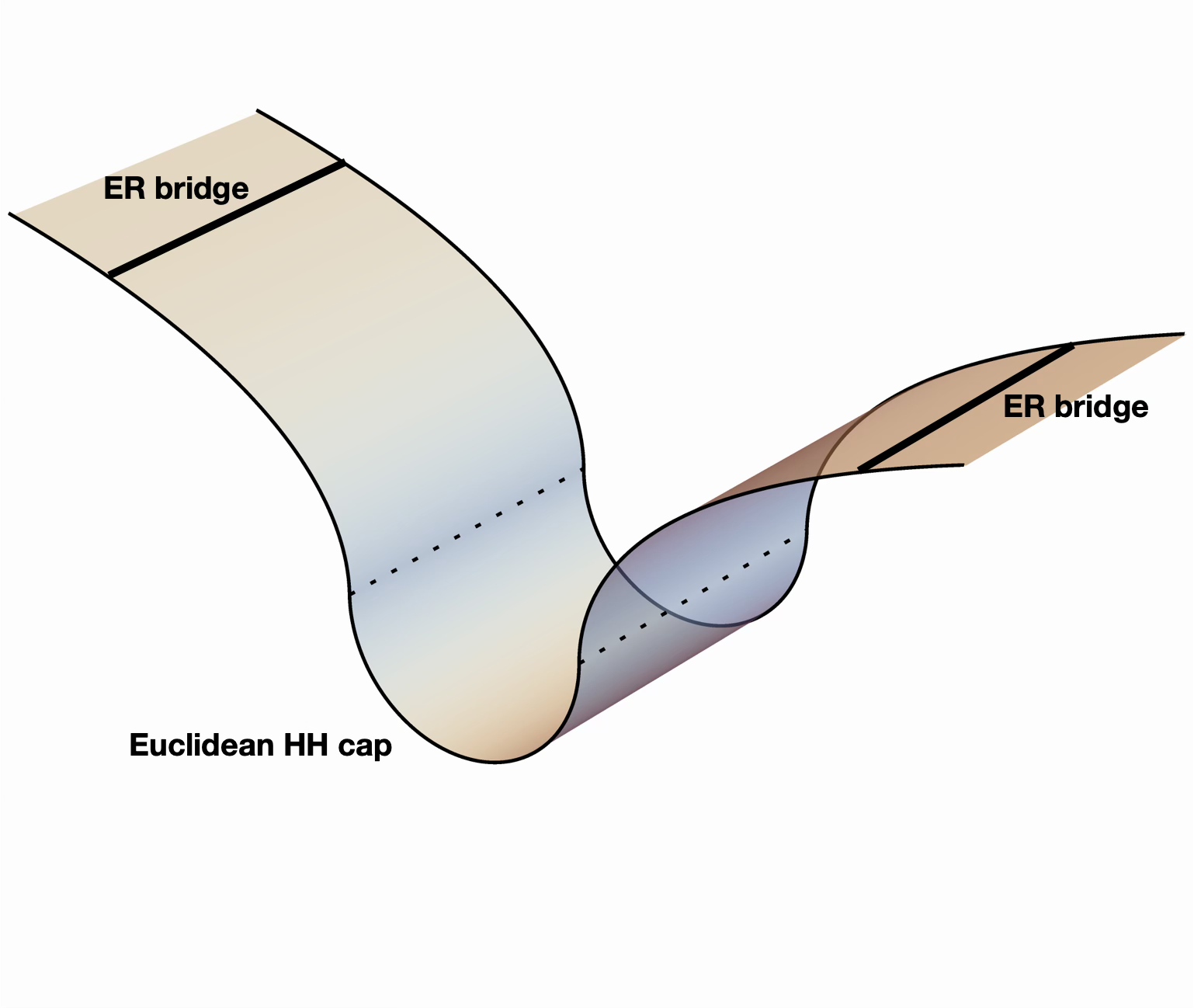}
\caption{Hartle--Hawking preparation of two ER bridges from a
two-boundary spacetime wormhole. The $S^1_\phi$ direction is
suppressed. The connected Euclidean half-geometry has two disconnected
ER components on its cutting surface, represented schematically by the
two curved strips. The corresponding path integral prepares a state in
$\mathcal H_{\mathrm{ER},1}\otimes\mathcal H_{\mathrm{ER},2}$, which
is generically entangled because the two ER components are prepared
together by the same connected half-geometry.}
\label{fig:ER_WH}
\end{figure}

We now contrast this with a two-boundary torus spacetime wormhole \cite{Cotler:2020ugk}. Its
bulk topology is that of a thickened torus,
\begin{equation}
    M_{\mathrm{WH}}
    \simeq
    T^2\times I
    \simeq
    I_{\rho}\times S^1_{\tau}\times S^1_{\phi},
    \qquad
    \partial M_{\mathrm{WH}}
    =
    T^2_L\sqcup T^2_R,
\end{equation}
where the ends of $I_{\rho}$ correspond to the two asymptotic torus
boundaries and
\begin{equation}
    T^2=S^1_{\tau}\times S^1_{\phi}.
\end{equation}
For $\tau\sim\tau+\beta$, the reflection $\tau\mapsto-\tau$ again has
fixed loci at $\tau=0$ and $\tau=\beta/2$. In contrast to the BTZ
case, however, the Euclidean-time circle does not contract anywhere
in the thickened torus. The two fixed surfaces therefore remain
disconnected:
\begin{equation}
    \Sigma_{\mathrm{cut}}
    =
    \Sigma_0\sqcup\Sigma_{\beta/2},
    \qquad
    \Sigma_0\simeq\Sigma_{\beta/2}
    \simeq I_{\rho}\times S^1_{\phi}.
\end{equation}
The reflection decomposition is consequently
\begin{equation}
    M_{\mathrm{WH}}
    =
    M_{\mathrm{WH},1/2}
    \cup_{\Sigma_{\mathrm{cut}}}
    \overline{M}_{\mathrm{WH},1/2},
\end{equation}
with the two components of the cutting surface evolving into two ER
bridges after Lorentzian continuation, as illustrated schematically
in Figure~\ref{fig:ER_WH}.

Applying the same analytic continuations as in
Eq.~\eqref{eq:BTZ-analytic-continuation} to the two reflection surfaces
of the thickened torus gives a different result. Since the
Euclidean-time circle does not contract, the two surfaces do not meet,
and their continuations remain two distinct ER spatial geometries. A
single half-wormhole therefore prepares a two-ER state,
\begin{equation}
    \vert\Psi_{\mathrm{WH},1/2}[B_L,B_R]\rangle
    \in
    \mathcal H_{\mathrm{ER}}^{(1)}
    \otimes
    \mathcal H_{\mathrm{ER}}^{(2)}.
\end{equation}
The reflected half prepares its conjugate, and gluing the two
half-geometries reconstructs the spacetime-wormhole amplitude:
\begin{equation}
    Z_{\mathrm{WH}}[B_L,B_R]
    =
    \langle
        \Psi_{\mathrm{WH},1/2}[B_L,B_R]
    \vert
        \Psi_{\mathrm{WH},1/2}[B_L,B_R]
    \rangle.
\end{equation}
The two ER components are thus both present in each half-geometry;
they are not themselves the ket and bra. Rather, the two
reflection-related half-geometries prepare the ket and bra of the same
two-ER state.

The distinction between the two constructions is therefore
\begin{equation}
    \begin{aligned}
        \partial_{\mathrm{cut}}M_{\mathrm{BTZ},1/2}
        &=
        \Sigma_{\mathrm{ER}},
        &
        \vert\Psi_{\mathrm{BTZ},1/2}\rangle
        &\in
        \mathcal H_{\mathrm{ER}},
        \\
        \partial_{\mathrm{cut}}M_{\mathrm{WH},1/2}
        &=
        \Sigma_{\mathrm{ER}}^{(1)}
        \sqcup
        \Sigma_{\mathrm{ER}}^{(2)},
        &
        \vert\Psi_{\mathrm{WH},1/2}\rangle
        &\in
        \mathcal H_{\mathrm{ER}}^{(1)}
        \otimes
        \mathcal H_{\mathrm{ER}}^{(2)}.
    \end{aligned}
\end{equation}
Both amplitudes arise by gluing two copies of a half-geometry, but
Euclidean BTZ is the norm of a one-ER state, whereas the
thickened-torus spacetime wormhole is the norm of a two-ER state.

The two ER components of the latter emerge together from a single
connected Euclidean half-geometry, which has the structure of a
Hartle--Hawking pair-creation geometry tensored with the remaining
spatial direction. It therefore naturally prepares a state of the
form
\begin{equation}
    \vert\Psi_{\mathrm{WH},1/2}\rangle
    =
    \sum_{A,B}K_{AB}\,
    \vert A\rangle_{\mathrm{ER},1}
    \otimes
    \vert B\rangle_{\mathrm{ER},2}.
    \label{eq:two-ER-state}
\end{equation}
Reflection symmetry identifies the Hilbert spaces of the two ER
components and allows their Schmidt bases to be chosen compatibly.
The state may therefore be written in the Schmidt spectral form
\begin{equation}
    \vert\Psi_{\mathrm{WH},1/2}\rangle
    =
    \int_{\Lambda} d\mu(\lambda)\,
    k_{\lambda}\,
    \vert\lambda\rangle_{\mathrm{ER},1}
    \otimes
    \vert\lambda\rangle_{\mathrm{ER},2},
    \qquad
    k_{\lambda}\geq0,
    \label{eq:two-ER-TFD-like}
\end{equation}
where $\Lambda$ denotes the Schmidt spectrum and $d\mu(\lambda)$
includes both discrete and continuous contributions. The state is
entangled whenever the support of $k_\lambda$ contains more than one
Schmidt sector.
The doubled-diagonal form in
Eq.~\eqref{eq:two-ER-TFD-like} has the same purification structure as
a thermofield-double state, and we refer to it as TFD-like in this
structural sense. Unlike the ordinary BTZ TFD, however, the Schmidt
coefficients $k_\lambda$ are not assumed to have a thermal form. 
The state naturally admits a third-quantized interpretation as the
counterpart of the ordinary BTZ TFD construction, with entire ER
geometries, rather than the left and right sectors of a single ER
geometry, forming the two entangled subsystems.
The distinction is summarized in Table~\ref{tab:btz-wormhole-tfd}.
The interpretation of the Schmidt
sectors $\lambda$ in terms of complete one-universe wavefunctions,
and the resulting effective ensemble description, will be developed
in Subsection~\ref{sec:ensembles}.

The same torus wormhole also admits the fixed-radial cut relevant to
third-quantized evolution,
\begin{equation}
    \Sigma_{\mathrm{rad}}
    =
    \{\rho=\rho_0\}
    \simeq
    S^1_{\tau}\times S^1_{\phi}
    =
    T^2.
\end{equation}
Cutting along this torus gives
\begin{equation}
    M_{\mathrm{WH}}
    =
    M_L\cup_{\Sigma_{\mathrm{rad}}}M_R,
\end{equation}
and expresses the wormhole amplitude as
\begin{equation}
    Z_{\mathrm{WH}}[B_L,B_R]
    =
    \int\mathcal D\gamma_{T^2}\,
    \Psi_L[B_L;\gamma_{T^2}]
    \Psi_R[\gamma_{T^2};B_R],
    \label{eq:radial-wormhole-gluing}
\end{equation}
where $\gamma_{T^2}$ denotes the gravitational data on the
intermediate torus. Upon expanding the two radial wavefunctions in a
basis of one-universe states, Eq.~\eqref{eq:radial-wormhole-gluing}
defines a bilinear two-universe kernel. In the third-quantized
description, this kernel determines the pair-correlated, or squeezed,
component of the spacetime-wormhole state, as discussed in Subsection~\ref{sec:STwormhole_states}.

The Euclidean-time and radial cuts therefore provide complementary
representations of the same codimension-zero geometry. The former
exhibits the half-wormhole as a Hartle--Hawking-type entangled state
of two complete ER geometries and is naturally adapted to its
Schmidt or TFD-like description. The latter is adapted to radial
evolution and expresses the same wormhole correlation through the
pair-correlated sector of the universe Fock space.

\newcolumntype{C}[1]{%
    >{\centering\arraybackslash}m{#1}%
}

\begin{table}[t]
    \centering
    \small
    \renewcommand{\arraystretch}{1.45}
    \setlength{\tabcolsep}{6pt}

    \begin{tabular}{
        |C{0.19\textwidth}
        |C{0.34\textwidth}
        |C{0.34\textwidth}|
    }
        \hline
        &
        \textbf{Euclidean BTZ}
        &
        \textbf{Torus spacetime wormhole}
        \\
        \hline

        \textbf{Euclidean topology}
        &
        \(D^2\times S^1\)
        &
        \(T^2\times I\)
        \\
        \hline

        \textbf{Half-geometry}
        &
        Half-solid-torus
        &
        Half of the thickened torus
        \\
        \hline

        \textbf{Cutting boundary}
        &
        One connected ER cylinder
        \newline
        \(\Sigma_{\mathrm{ER}}\simeq I\times S^1\)
        &
        Two ER cylinders
        \newline
        \(\Sigma_{\mathrm{ER}}^{(1)}
        \sqcup\Sigma_{\mathrm{ER}}^{(2)}\)
        \\
        \hline

        \textbf{State space}
        &
        \(\mathcal H_{\mathrm{ER}}
        \simeq\mathcal H_L\otimes\mathcal H_R\)
        &
        \(\mathcal H_{\mathrm{ER}}^{(1)}
        \otimes\mathcal H_{\mathrm{ER}}^{(2)}\)
        \\
        \hline

        \textbf{Prepared state}
        &
        Ordinary TFD
        \newline
        \(\displaystyle
        \vert\mathrm{TFD}\rangle
        =
        \sum_n e^{-\beta E_n/2}
        \vert n\rangle_L\vert n\rangle_R\)
        &
        Third-quantized TFD-like state
        \newline
        \(\displaystyle
        \vert\Psi_{\mathrm{WH},1/2}\rangle
        =
        \sum_{A,B}K_{AB}
        \vert A\rangle_1\vert B\rangle_2\)
        \\
        \hline

        \textbf{Entangled sectors}
        &
        Left and right sectors within one ER geometry
        &
        Two complete ER geometries
        \\
        \hline

        \textbf{Gluing result}
        &
        \(\displaystyle
        Z_{\mathrm{BTZ}}
        =
        \langle\mathrm{TFD}
        \vert\mathrm{TFD}\rangle\)
        &
        \(\displaystyle
        Z_{\mathrm{WH}}[B_L,B_R]
        =
        \langle\Psi_{\mathrm{WH},1/2}
        \vert\Psi_{\mathrm{WH},1/2}\rangle\)
        \\
        \hline

        \textbf{Interpretation}
        &
        Entanglement in the one-universe sector
        &
        Entanglement in the multi-universe sector
        \\
        \hline
    \end{tabular}

    \caption{Comparison between the Hartle--Hawking preparation of the
    BTZ thermofield-double state and the two-ER state prepared by half
    of a torus spacetime wormhole.}
    \label{tab:btz-wormhole-tfd}
\end{table}

The two Euclidean fillings are topologically distinct:
\begin{equation}
    \begin{aligned}
        M_{\mathrm{BTZ}}^{(\mathrm E)}
        &\simeq D^2\times S^1,
        &
        \partial M_{\mathrm{BTZ}}^{(\mathrm E)}
        &=T^2,
        \\
        M_{\mathrm{WH}}
        &\simeq T^2\times I,
        &
        \partial M_{\mathrm{WH}}
        &=T^2_L\sqcup T^2_R.
    \end{aligned}
\end{equation}
Correspondingly, their reflection halves prepare states in different
sectors. The BTZ half-solid-torus prepares a one-ER state in the
one-universe Hilbert space, whereas half of the connected
two-boundary torus wormhole prepares an entangled two-ER state in
$\operatorname{Sym}^{2}\mathcal H_{\mathrm U}$. More generally, a
Euclidean half-geometry with $N$ connected ER components prepares a
state in the $N$-universe sector of the universe Fock space. Cutting
spacetime-wormhole geometries thus directly exposes the multi-universe
state space for which third quantization provides the natural quantum
description.

\subsection{Spacetime-Wormhole Correlations}
\label{sec:STwormhole_states}

Having identified the multi-universe state obtained by cutting a
spacetime wormhole, we now describe its third-quantized construction.
The absence of an exposed incoming boundary does not mean that the
geometry is prepared from the Fock vacuum. Rather, its incoming leg is
terminated by a coherent-state cap, and topology-changing interactions
acting on this background generate correlations among the outgoing
universes.

To recover the ordinary fixed-topology description, we take the
initial third-quantized state to be a coherent state
$\vert\Omega\rangle$, defined by
\begin{equation}
    a_A\vert\Omega\rangle
    =
    \alpha_A\vert\Omega\rangle.
\end{equation}
Its one-point function is
\begin{equation}
    \langle\Omega|
    \widehat\Psi(q,T)
    |\Omega\rangle
    =
    \sum_A
    \alpha_A\psi_A(q,T)
    =
    \Psi_{\mathrm{norm}}(q,T).
\end{equation}
Thus, specializing
Eq.~\eqref{eq:full-wavefunction-expectation} to
$\vert\Phi\rangle=\vert\Omega\rangle$ recovers the conventional
one-universe wavefunction after the fixed non-normalizable
contribution is restored. The coherent state therefore embeds the
fixed-topology description into the universe Fock space.

The coherent state $\vert\Omega\rangle$ plays a role analogous to that
of a Coleman $\alpha$-state
\cite{Coleman:1988cy,Coleman:1988tj}: it assigns definite expectation
values to the universe field and thereby selects a fixed one-universe
wavefunction. The analogy is not exact. Coleman $\alpha$-states are
conventionally defined as eigenstates of commuting Hermitian
baby-universe operators, whereas $\vert\Omega\rangle$ is an eigenstate
of the annihilation operators. The latter is natural in the reduced
phase-space formulation, in which the physical wavefunction is
promoted directly to an annihilation field.

Throughout this subsection, we work in the interaction picture. The
free radial evolution is carried by the mode functions appearing in
$\widehat\Psi(q,T)$, and hence by the corresponding boundary operator
$\widehat Z(\tau;T)$. Changes in topology and universe number are
generated by
\begin{equation}
    U_I(T_{\mathrm f},T_0)
    =
    \mathcal T
    \exp\!\left[
        -i\int_{T_0}^{T_{\mathrm f}}dT\,
        \widehat H_{\mathrm{int}}^I(T)
    \right],
    \label{eq:topology-changing-evolution}
\end{equation}
where $T_0$ and $T_{\mathrm f}$ denote the initial and final endpoints
of the radial evolution. The resulting spacetime-wormhole state is
\begin{equation}
    \vert\Psi_{\mathrm{WH}}\rangle_I
    :=
    U_I(T_{\mathrm f},T_0)\vert\Omega\rangle.
    \label{eq:WH-state}
\end{equation}
It contains all time-ordered splitting and joining histories generated
between the two radial endpoints.

To expose its multi-universe content, we keep the coherent cap
$\vert\Omega\rangle$ unexpanded and normal-order the evolution
operator before allowing it to act on the cap. All annihilation
operators can then be moved to the right and replaced by their
coherent-state eigenvalues. The state consequently takes the form
\begin{equation}
    \vert\Psi_{\mathrm{WH}}\rangle_I
    =
    \sum_{N=0}^{\infty}
    \frac{1}{N!}
    \sum_{A_1,\ldots,A_N}
    \mathcal C_{A_1\cdots A_N}\,
    a_{A_1}^\dagger\cdots a_{A_N}^\dagger
    \vert\Omega\rangle.
    \label{eq:WH-leg-expansion}
\end{equation}
Here and below, sums over mode labels include the appropriate spectral
integrals. In Eq.~\eqref{eq:WH-leg-expansion}, $N$ counts the explicit
outgoing universe legs created above the coherent-state cap. It should
therefore be understood as a normal-ordered leg expansion relative to
the unexpanded cap, rather than as an expansion of
$\vert\Omega\rangle$ into eigenstates of the Fock-space number
operator.

The coefficients $\mathcal C_{A_1\cdots A_N}$ receive contributions
from all interaction histories with $N$ outgoing legs. Schematically,
if the normal-ordered evolution operator is written as
\begin{align}
    U_I
    =
    \sum_{N,M\geq0}
    \frac{1}{N!\,M!}\,
    \mathcal U_{A_1\cdots A_N}{}^{B_1\cdots B_M}\,
    a_{A_1}^\dagger\cdots a_{A_N}^\dagger
    a_{B_1}\cdots a_{B_M},
\end{align}
then
\begin{equation}
    \mathcal C_{A_1\cdots A_N}
    =
    \sum_{M=0}^{\infty}
    \frac{1}{M!}
    \sum_{B_1,\ldots,B_M}
    \mathcal U_{A_1\cdots A_N}{}^{B_1\cdots B_M}
    \alpha_{B_1}\cdots\alpha_{B_M}.
    \label{eq:WH-leg-coefficients}
\end{equation}
The incoming legs of the normal-ordered interaction histories are thus
absorbed by the coherent cap, while the remaining creation operators
represent the outgoing universes.

By construction, every topology-changing vertex in
$\widehat H_{\mathrm{int}}^I(T)$ is accompanied by its
Hermitian-conjugate process. The interaction Hamiltonian is therefore
Hermitian,
\begin{equation}
    \widehat H_{\mathrm{int}}^I(T)
    =
    \bigl(
        \widehat H_{\mathrm{int}}^I(T)
    \bigr)^\dagger,
\end{equation}
and the interaction-picture evolution operator is unitary:
\begin{equation}
    U_I^\dagger(T_{\mathrm f},T_0)
    U_I(T_{\mathrm f},T_0)
    =
    \mathbf 1.
\end{equation}
Consequently, for a normalized coherent state,
\begin{equation}
    {}_I\langle\Psi_{\mathrm{WH}}
    \vert\Psi_{\mathrm{WH}}\rangle_I
    =
    \langle\Omega|
    U_I^\dagger(T_{\mathrm f},T_0)
    U_I(T_{\mathrm f},T_0)
    |\Omega\rangle
    =
    1.
\end{equation}
The connected two-wavefunction correlator in this state is
\begin{align}
    \mathcal W_{\mathrm{WH}}(1,2)
    :={}
    {}_I\langle\Psi_{\mathrm{WH}}|
    \widehat\Psi(1)\widehat\Psi(2)
    |\Psi_{\mathrm{WH}}\rangle_I
    -
    {}_I\langle\Psi_{\mathrm{WH}}|
    \widehat\Psi(1)
    |\Psi_{\mathrm{WH}}\rangle_I
    \,
    {}_I\langle\Psi_{\mathrm{WH}}|
    \widehat\Psi(2)
    |\Psi_{\mathrm{WH}}\rangle_I,
    \label{eq:WH-connected-wavefunction-correlator}
\end{align}
where
$\widehat\Psi(i)\equiv\widehat\Psi(q_i,T_0)$ and $T_0$ denotes the
asymptotic boundary. Equivalently,
\begin{equation}
    \mathcal W_{\mathrm{WH}}(1,2)
    =
    \langle\Omega|
    U_I^\dagger\,
    \widehat\Psi(1)\widehat\Psi(2)\,
    U_I
    |\Omega\rangle_{\mathrm{conn}},
    \label{eq:WH-connected-coherent-expectation}
\end{equation}
with the arguments $(T_{\mathrm f},T_0)$ of $U_I$ left implicit. The
inserted fields carry the free radial evolution and are evaluated at
the asymptotic boundary, while $U_I$ contains the topology-changing
evolution through the bulk.

In the mode basis, the corresponding connected pair matrix is
\begin{align}
    C_{AB}^{\mathrm{WH}}
    :={}
    {}_I\langle\Psi_{\mathrm{WH}}|
    a_Aa_B
    |\Psi_{\mathrm{WH}}\rangle_I
    -
    {}_I\langle\Psi_{\mathrm{WH}}|
    a_A
    |\Psi_{\mathrm{WH}}\rangle_I
    \,
    {}_I\langle\Psi_{\mathrm{WH}}|
    a_B
    |\Psi_{\mathrm{WH}}\rangle_I.
    \label{eq:WH-connected-pair-matrix}
\end{align}
Together with the freely evolved mode functions, this matrix determines
the connected wavefunction correlator:
\begin{equation}
    \mathcal W_{\mathrm{WH}}(1,2)
    =
    \sum_{A,B}
    \psi_A(q_1,T_0)\,
    \psi_B(q_2,T_0)\,
    C_{AB}^{\mathrm{WH}},
    \label{eq:WH-wavefunction-mode-expansion}
\end{equation}
where the sums include the appropriate continuous spectral integrals.

In the torus example, let
\begin{equation}
    \widehat Z_\pm
    :=
    \widehat Z(\beta\pm it;T_0),
    \label{eq:complexified-partition-operators}
\end{equation}
where $\beta\pm it$ denotes the complexified inverse temperature of
the boundary theory. The spacetime-wormhole contribution to the
spectral form factor is the connected cumulant
\begin{align}
    K_{\mathrm{WH}}(\beta,t)
    :={}
    {}_I\langle\Psi_{\mathrm{WH}}|
    \widehat Z_+\widehat Z_-
    |\Psi_{\mathrm{WH}}\rangle_I
    -
    {}_I\langle\Psi_{\mathrm{WH}}|
    \widehat Z_+
    |\Psi_{\mathrm{WH}}\rangle_I
    \,
    {}_I\langle\Psi_{\mathrm{WH}}|
    \widehat Z_-
    |\Psi_{\mathrm{WH}}\rangle_I.
    \label{eq:WH-connected-SFF}
\end{align}
Equivalently,
\begin{equation}
    K_{\mathrm{WH}}(\beta,t)
    =
    \langle\Omega|
    U_I^\dagger\,
    \widehat Z_+\widehat Z_-\,
    U_I
    |\Omega\rangle_{\mathrm{conn}}.
\end{equation}
Here $T$ is the radial evolution parameter and should be distinguished
from the field-theory time $t$ entering the complexified inverse
temperature. The one-point functions in the disconnected subtraction
are the complete interaction-dressed one-boundary amplitudes. Their
product removes contributions in which the two marked boundary
insertions belong to distinct connected components.

The perturbative structure of the correlator may be displayed
equivalently by moving $U_I$ onto the inserted operators. For the pair
operator,
\begin{align}
    U_I^\dagger a_Aa_B U_I
    ={}&
    a_Aa_B
    +
    i\int_{T_0}^{T_{\mathrm f}}dT_1\,
    \bigl[
        \widehat H_{\mathrm{int}}^I(T_1),
        a_Aa_B
    \bigr]
    \nonumber\\
    &+
    i^2
    \int_{T_0}^{T_{\mathrm f}}dT_1
    \int_{T_0}^{T_1}dT_2\,
    \bigl[
        \widehat H_{\mathrm{int}}^I(T_2),
        \bigl[
            \widehat H_{\mathrm{int}}^I(T_1),
            a_Aa_B
        \bigr]
    \bigr]
    +\cdots .
    \label{eq:WH-nested-commutators}
\end{align}
No factorials appear because the integration region is already time
ordered. The higher nested commutators incorporate repeated splitting
and joining processes and all intermediate universe-number
configurations.

The leading connected contribution follows from the elementary
splitting vertex. With the convention
\begin{equation}
    \widehat H_{\mathrm{split}}^I(T)
    =
    \frac{1}{2}
    \sum_{M,N,C}
    g_{MN}{}^C(T)\,
    a_M^\dagger a_N^\dagger a_C,
    \qquad
    g_{MN}{}^C=g_{NM}{}^C,
    \label{eq:elementary-splitting-vertex}
\end{equation}
and
$\widehat H_{\mathrm{join}}^I
=
(\widehat H_{\mathrm{split}}^I)^\dagger$, one finds
\begin{align}
    \bigl[
        \widehat H_{\mathrm{split}}^I,
        a_Aa_B
    \bigr]
    ={}
    -
    \sum_{M,C}
    g_{AM}{}^C\,
    a_M^\dagger a_Ca_B
    -
    \sum_{M,C}
    g_{BM}{}^C\,
    a_M^\dagger a_Ca_A
    -
    \sum_C
    g_{AB}{}^C\,a_C.
    \label{eq:splitting-aa-commutator}
\end{align}
The first two terms arise from single contractions. Their
coherent-state expectation values have the form
$\alpha^*\alpha\alpha$ and contribute to the
interaction-induced variation of the product of one-point functions.
They are therefore removed when the factorized contribution is
subtracted in Eq.~\eqref{eq:WH-connected-pair-matrix}. The final term
comes from contracting both annihilation operators with the two
creation operators at the same splitting vertex and provides the
leading connected source:
\begin{equation}
    C_{AB}^{\mathrm{WH}}
    =
    -i
    \int_{T_0}^{T_{\mathrm f}}dT\,
    \sum_C
    g_{AB}{}^C(T)\,
    \alpha_C
    +
    O\!\left(
        H_{\mathrm{int}}^2
    \right).
    \label{eq:WH-leading-pair-matrix}
\end{equation}
The same expression gives the leading contribution to the two-leg
amplitude $\mathcal C_{AB}$ in
Eq.~\eqref{eq:WH-leg-expansion}.

\begin{figure}[!h]
    \vspace{.2cm}
    \centering
    \includegraphics[height=1.6in]{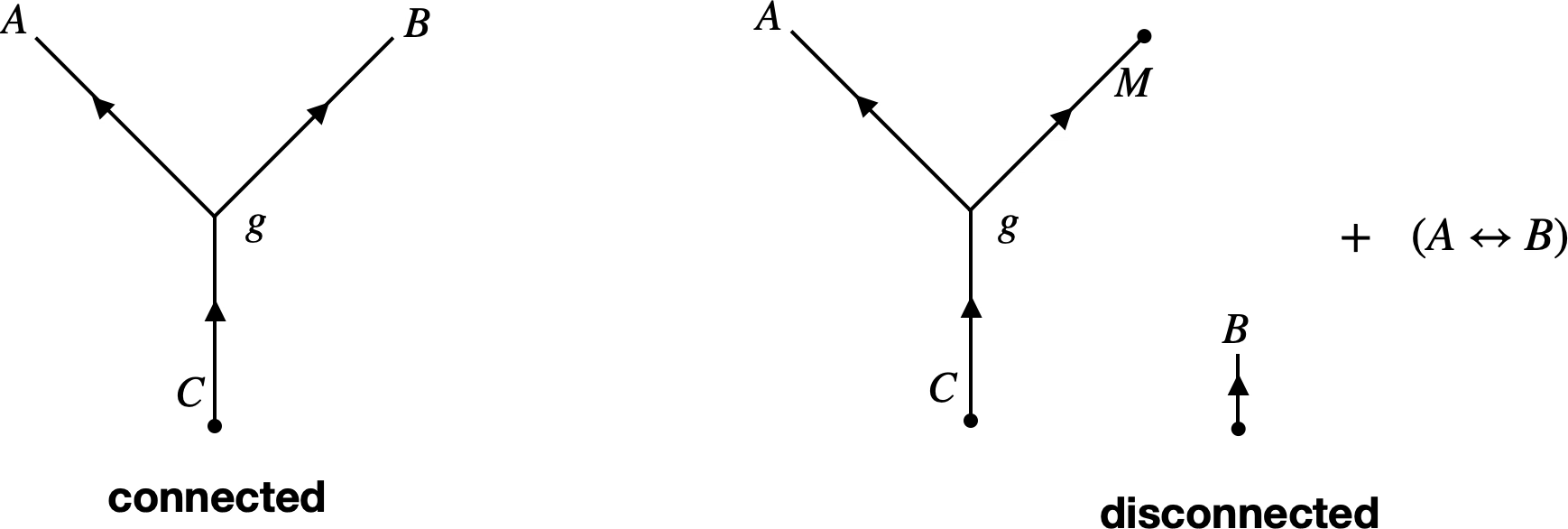}
    \caption{Coherent-state contractions in
    Eq.~\eqref{eq:splitting-aa-commutator}. Open ends represent exposed
    boundaries, and black dots represent coherent-state caps. A mode
    label at an open end identifies the exposed boundary, whereas a
    label beside a capped line indicates that the corresponding
    oscillator has been replaced by its coherent-state eigenvalue.
    Arrows distinguish creation and annihilation legs. The left diagram
    represents the connected double-contraction term
    $g_{AB}{}^C\alpha_C$. The right diagram represents the
    single-contraction contribution
    $g_{AM}{}^C\alpha_M^*\alpha_C\alpha_B$, together with
    $A\leftrightarrow B$, which factorizes into two one-boundary
    components.}
    \label{fig:diagram1}
\end{figure}

\begin{figure}[!h]
    \vspace{.2cm}
    \centering
    \includegraphics[height=2.0in]{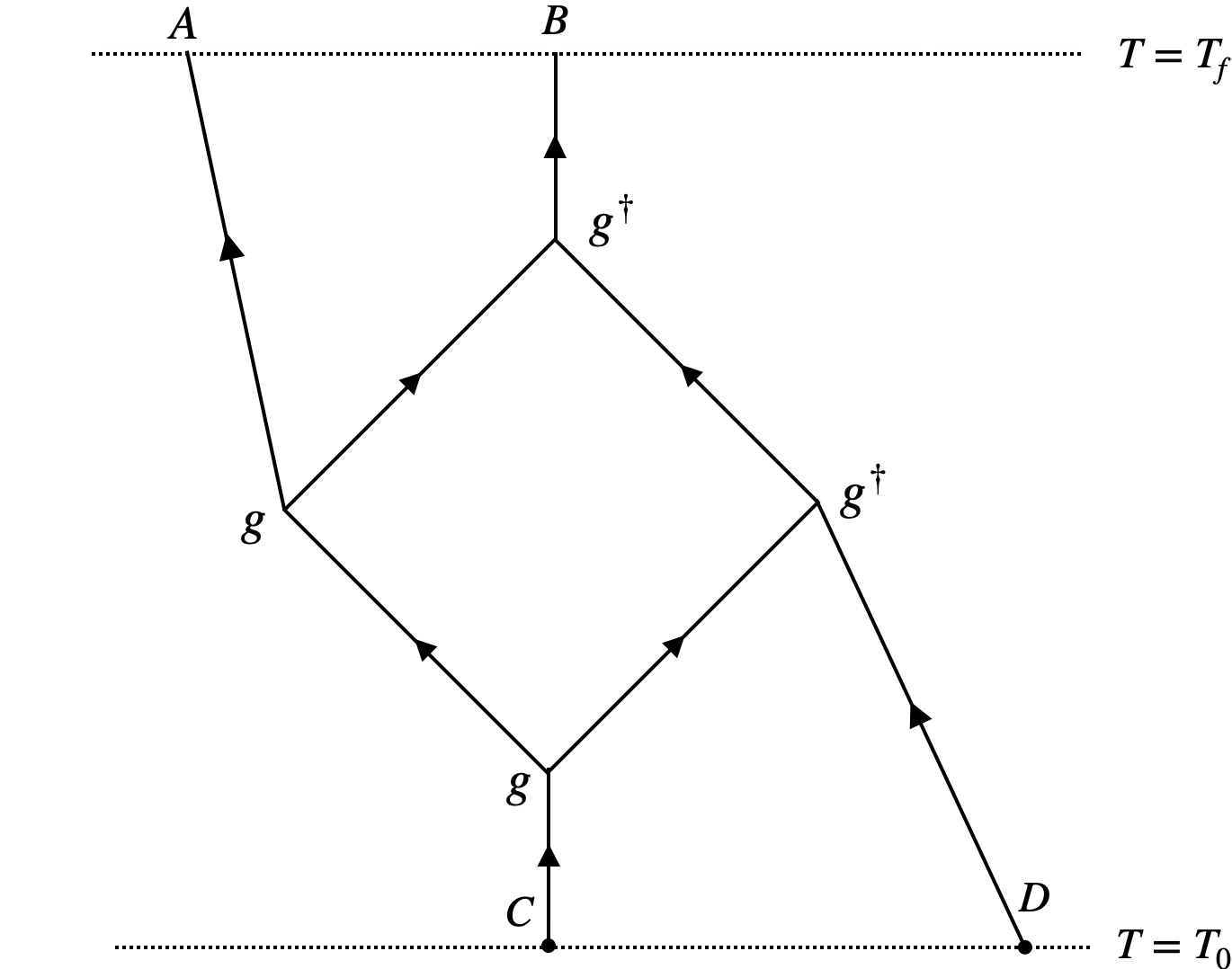}
    \caption{A representative higher-order connected contribution in
    radially ordered perturbation theory. The legs $C$ and $D$
    terminate on the coherent-state cap at $T=T_0$, while $A$ and $B$
    remain exposed at $T=T_{\mathrm f}$. Two splitting vertices $g$
    and two joining vertices $g^\dagger$ form an internal universe
    loop.}
    \label{fig:loop}
\end{figure}

The corresponding contractions are represented in
Figure~\ref{fig:diagram1}. The connected diagram depicts the double
contraction that produces Eq.~\eqref{eq:WH-leading-pair-matrix},
whereas the disconnected diagram depicts the single-contraction terms
removed upon forming the connected correlator. The joining vertex has
no analogous double contraction at this order, although it contributes
through higher nested commutators. A representative higher-order
connected contribution involving both splitting and joining vertices
is shown in Figure~\ref{fig:loop}.

\begin{figure}[!ht]
    \vspace{.2cm}
    \centering
    \includegraphics[height=2.5in]{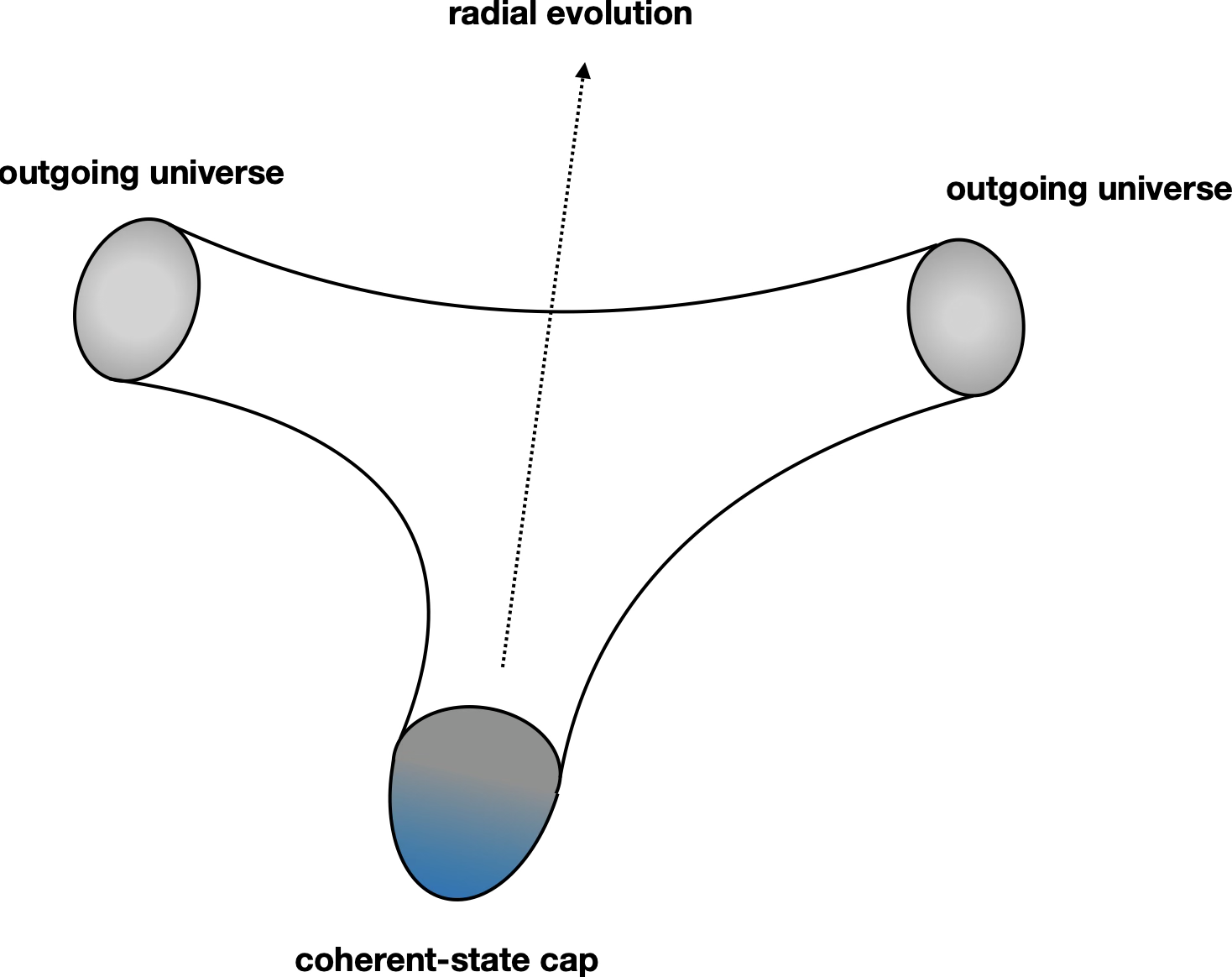}
    \caption{Geometric thickening of the leading connected skeleton in
    Figure~\ref{fig:diagram1}. Each universe line becomes a tube and
    the splitting vertex becomes a pair-of-pants cobordism. The two
    upper ends are exposed outgoing boundaries, while the incoming leg
    terminates on the coherent-state cap. The arrow indicates the direction of radial evolution.}
    \label{fig:WH_state}
\end{figure}

The skeleton diagrams admit a direct geometric interpretation.
Thickening each universe line into a tube and each trivalent vertex
into a pair-of-pants cobordism turns the leading connected diagram in
Figure~\ref{fig:diagram1} into the capped two-boundary geometry shown
in Figure~\ref{fig:WH_state}. The coherent state is kept unexpanded:
after normal ordering, annihilation operators act on the cap and are
replaced by their eigenvalues, while the remaining creation operators
give the exposed outgoing legs.

By contrast, the disconnected diagram thickens into a disjoint union
of two one-boundary fillings. The component containing the interaction
vertex gives a dressed one-point amplitude for one marked boundary,
while the separate capped line gives the one-point amplitude for the
other. Thus, the algebraic subtraction of the factorized contribution
precisely matches the removal of disconnected bulk geometries.
Similarly, thickening Figure~\ref{fig:loop} produces a connected
two-boundary geometry with an internal handle.

The cancellation of factorized terms persists beyond leading order.
Because the wormhole contribution is defined as a connected cumulant,
the linked-cluster expansion removes all terms that decompose into
products of lower-point expectation values. Higher-order connected
terms nevertheless receive contributions from splitting, joining, and
more general vertices through histories involving arbitrary
intermediate universe-number configurations. Terms proportional to
the identity do not enter the nested-commutator expansion because they
commute with the inserted operators.

At leading order, $C_{AB}^{\mathrm{WH}}$ also serves as the coefficient
matrix of the two outgoing legs of the capped splitting geometry. If
the legs are assigned distinguishable one-universe factors, the
resulting pair is entangled whenever this matrix has rank greater than
one. Beyond leading order, however, the full state contains more
general topology-changing histories, and its entanglement cannot be
inferred from the leading pair matrix alone.

This entanglement provides a connection with the Euclidean-time cut
discussed in Subsection~\ref{sec:ER}. That cut represents the same
codimension-zero spacetime wormhole as a TFD-like state of two ER
geometries. The radial and Euclidean-time cuts therefore give
complementary representations of the two-universe correlations carried
by the wormhole.

More precisely, let $C^{(2)}$ denote the coefficient matrix of the
two-universe component in the radial description and let $K$ denote
the coefficient matrix of the two-ER state obtained from the
Euclidean-time cut. At leading order, $C^{(2)}$ is determined by
Eq.~\eqref{eq:WH-leading-pair-matrix}. If the change of slicing acts
through invertible maps $U_1$ and $U_2$ on the two one-universe
factors, then
\begin{equation}
    K
    =
    U_1 C^{(2)} U_2^{\mathrm T}.
\end{equation}
It follows that
\begin{equation}
    \operatorname{rank}K
    =
    \operatorname{rank}C^{(2)},
\end{equation}
so separability is preserved under the change of representation.
Constructing $U_1$ and $U_2$ explicitly requires control over the
change between radial and Euclidean-time slicings. Subject to this
qualification, the TFD-like two-ER state and the pair-correlated radial
state encode the same underlying wormhole correlations.

\subsection{Sewing Consistency and the Third-Quantized Bootstrap}
\label{sec:third-quantized-bootstrap}

The preceding construction assumes the existence of splitting and
joining vertices, but does not yet determine their coupling tensors.
A natural set of constraints follows from gravitational sewing. The
same connected cobordism may admit different decompositions into
elementary topology-changing processes, and the resulting amplitude
must be independent of this auxiliary choice of decomposition. This
suggests a third-quantized bootstrap in which the interaction vertices
are constrained by the consistency of all possible cutting and gluing
channels.

Consider a connected amplitude with four exposed universe legs. One
possible decomposition joins $A$ and $B$ into an intermediate
universe $E$, which subsequently splits into $C$ and $D$. In the
interaction picture, the corresponding exchange contribution is
\begin{equation}
    \mathcal E_{ABCD}^{(s)}
    :=
    \int_{T_0}^{T_{\mathrm f}}dT_2
    \int_{T_0}^{T_2}dT_1
    \sum_E
    g_{CD}{}^E(T_2)\,
    g^{AB}{}_{E}(T_1).
    \label{eq:s-channel-exchange}
\end{equation}
The sum over $E$ includes integration over any continuous part of the
one-universe spectrum. The ordered radial integrals and the
$T$-dependence of the interaction-picture couplings incorporate the
free propagation of the intermediate universe; no additional
propagator is required.

An alternative decomposition joins $A$ and $C$ and gives
\begin{equation}
    \mathcal E_{ABCD}^{(t)}
    :=
    \int_{T_0}^{T_{\mathrm f}}dT_2
    \int_{T_0}^{T_2}dT_1
    \sum_E
    g_{BD}{}^E(T_2)\,
    g^{AC}{}_{E}(T_1).
    \label{eq:t-channel-exchange}
\end{equation}
The complete four-leg amplitude may also contain a four-universe
contact contribution $\mathcal A_{ABCD}^{(4)}$. Since this term is
intrinsic to the four-leg amplitude and does not depend on the
exchange channel, the two representations take the form
\begin{equation}
    \mathcal A_{ABCD}^{(s)}
    =
    \mathcal E_{ABCD}^{(s)}
    +
    \mathcal A_{ABCD}^{(4)},
    \qquad
    \mathcal A_{ABCD}^{(t)}
    =
    \mathcal E_{ABCD}^{(t)}
    +
    \mathcal A_{ABCD}^{(4)}.
    \label{eq:four-leg-channel-decompositions}
\end{equation}
Sewing consistency requires the complete amplitudes to agree.
Because the contact contribution is common to the two
representations, it drops out of this condition, leaving
\begin{equation}
    \mathcal E_{ABCD}^{(s)}
    =
    \mathcal E_{ABCD}^{(t)}.
    \label{eq:third-quantized-crossing}
\end{equation}
A sufficient, though generally stronger than necessary, condition for
Eq.~\eqref{eq:third-quantized-crossing} is equality of the
radially ordered integrands:
\begin{equation}
    \sum_E
    g_{CD}{}^E(T_2)\,
    g^{AB}{}_{E}(T_1)
    =
    \sum_E
    g_{BD}{}^E(T_2)\,
    g^{AC}{}_{E}(T_1),
    \qquad
    T_0\leq T_1\leq T_2\leq T_{\mathrm f}.
    \label{eq:local-third-quantized-crossing}
\end{equation}
This local sewing condition guarantees equality of the integrated
exchange amplitudes. The converse need not hold, since distinct
integrands may give the same radially integrated amplitude.

Equation~\eqref{eq:third-quantized-crossing} is analogous to crossing
symmetry of a CFT four-point function, but its origin is different.
CFT crossing expresses associativity of the operator product expansion
within a fixed boundary theory. The present condition expresses the
independence of a gravitational cobordism amplitude from the choice of
exchange channel used in its decomposition into elementary splitting
and joining processes. The two conditions must be compatible whenever
the universe legs admit a boundary-CFT realization, but they constrain
different structures.

The pointwise relation
\eqref{eq:local-third-quantized-crossing} is a
local-in-radial-time sewing condition for the cubic vertices. It is a
nontrivial functional identity involving sums or integrals over
complete intermediate one-universe states, with their radial, moduli,
and topological dependence encoded in the interaction-picture
couplings. Even when it holds, higher $m\to n$ vertices may still be
required to describe regions of cobordism moduli space not generated
by cubic sewing and to avoid overcounting. The bootstrap therefore
constrains both the cubic couplings and the higher vertices of the
third-quantized Hamiltonian.

A complementary set of constraints follows from the
Schwinger--Dyson hierarchy of the third-quantized theory. For an
interaction-picture operator $\widehat{\mathcal O}$ inserted at a
fixed radial position, define
\begin{equation}
    \langle\widehat{\mathcal O}\rangle_{\Omega;T_{\mathrm f},T_0}
    :=
    \langle\Omega|
    U_I^\dagger(T_{\mathrm f},T_0)\,
    \widehat{\mathcal O}\,
    U_I(T_{\mathrm f},T_0)
    |\Omega\rangle.
    \label{eq:evolved-coherent-expectation}
\end{equation}
Differentiating this expectation value with respect to the final
endpoint of the interaction evolution gives
\begin{equation}
    i\frac{\partial}{\partial T_{\mathrm f}}
    \langle\widehat{\mathcal O}\rangle_{\Omega;T_{\mathrm f},T_0}
    =
    \left\langle
        \left[
            \widehat{\mathcal O},
            \widehat H_{\mathrm{int}}^I(T_{\mathrm f})
        \right]
    \right\rangle_{\Omega;T_{\mathrm f},T_0}.
    \label{eq:third-quantized-Schwinger-Dyson}
\end{equation}
Equation~\eqref{eq:third-quantized-Schwinger-Dyson} follows by
differentiating the interaction-picture evolution operators with
respect to their final endpoint.
For
$\widehat{\mathcal O}
=
\widehat\Psi(q_1,T_0)\widehat\Psi(q_2,T_0)$, the field insertions are
held fixed at the asymptotic boundary, so there is no explicit
$T_{\mathrm f}$ derivative of $\widehat{\mathcal O}$. 
For a cubic splitting and joining Hamiltonian,
Eq.~\eqref{eq:third-quantized-Schwinger-Dyson} relates the evolution
of the connected two-wavefunction correlator to three-wavefunction
correlators and coherent one-point functions. In the mode basis, the
same relation follows by taking
$\widehat{\mathcal O}=a_Aa_B$. When a multi-universe amplitude is
known only after the radial evolution has been completed, applying
this relation requires keeping the final endpoint
$T_{\mathrm f}$ finite and considering the corresponding family of
partially evolved amplitudes. The two-universe amplitude can then be
used as input to constrain the three-universe amplitude. Continuing
the hierarchy relates three-universe correlators to four-universe
correlators, and so forth.
These Schwinger--Dyson relations provide dynamical constraints across
different universe-number sectors, while the sewing conditions require
consistency among different decompositions of a given multi-boundary
amplitude.

In Section~\ref{sec:AdS3}, we use the known two-torus wormhole
amplitude as the leading two-universe input. In addition to
constraining a compatible splitting interaction, its finite-endpoint
extension may serve as the starting datum for the third-quantized
Schwinger--Dyson hierarchy.

\subsection{Replica-Wormhole Correlations}
\label{sec:replica_wormholes}

Replica wormholes are spacetime-wormhole contributions selected by the
boundary sewing conditions of the replica construction. To compute the
$n$th replica moment, one introduces $n$ labeled copies of the boundary
data and cyclically sews their subsystem-$A$ factors. In the
third-quantized description, these boundary conditions probe an
$n$-point universe-field correlation in the spacetime-wormhole state
introduced in Eq.~\eqref{eq:WH-state},
\begin{equation}
    \vert\Psi_{\mathrm{WH}}\rangle_I
    =
    U_I(T_{\mathrm f},T_0)\vert\Omega\rangle .
\end{equation}
Replica wormholes therefore do not require a separate
third-quantized state. They are particular connectivity sectors of
correlation functions evaluated in the complete topology-changing
state.

The relevant $n$-leg contribution need not be generated by a single
$1\to n$ vertex. It contains all radially ordered histories compatible
with the prescribed replica sewing, including sequences such as
$1\to2\to\cdots\to n$, as well as histories involving joining vertices
and intermediate sectors with different universe number. The replica
boundary condition selects the appropriate combination of these
histories from the full third-quantized evolution.

We first recall the replica construction when the boundary data admit a
microscopic quantum-theory realization. Assuming the relevant
factorization,
\begin{equation}
    \mathcal H_{\partial}
    =
    \mathcal H_A\otimes\mathcal H_{\bar A},
\end{equation}
the cyclic permutation $\mathbb V_A^{(n)}$ acts on the $A$ factors of
$n$ labeled copies and gives the standard replica identity
\begin{equation}
    \operatorname{Tr}\rho_A^n
    =
    \operatorname{Tr}
    \left[
        \rho^{\otimes n}\mathbb V_A^{(n)}
    \right].
\end{equation}
This microscopic construction specifies the boundary sewing rule that
must be implemented gravitationally.

The replica construction requires more information than an ordinary
one-universe wavefunction. Merely marking the entangling surface
$\partial A$ does not change the state. Before the replicas are sewn,
the path integral must instead be cut open along $A$, producing a
wavefunctional
\begin{equation}
    \Psi_{\lambda,A}
    \left[
        q;
        \xi_A^+,\xi_A^-;
        T
    \right]
    \label{eq:cut-open-universe-wavefunction}
\end{equation}
with independent data $\xi_A^\pm$ on the two sides of the cut. The
asymptotic boundary map extends to these cut-open wavefunctionals as
\begin{equation}
    \mathcal B_A:
    \Psi_{\lambda,A}
    \left[
        q;
        \xi_A^+,\xi_A^-;
        T
    \right]
    \longmapsto
    Z_{\lambda,A}
    \left[
        q;
        \xi_A^+,\xi_A^-
    \right]
    :=
    \Psi_{\lambda,A}
    \left[
        q;
        \xi_A^+,\xi_A^-;
        T_0
    \right].
    \label{eq:cut-open-boundary-value-map}
\end{equation}
Here $Z_{\lambda,A}$ is a boundary functional with open-cut data. 
When the one-universe wavefunction admits a consistent microscopic
boundary-theory interpretation, the cut-open path integral prepares
the corresponding density matrix. More generally, it is simply a boundary
functional of the two sets of open-cut data. A generic gravitational boundary functional need not admit this
density-matrix interpretation.

We denote by $\mathfrak S_A^{(n)}$ the multilinear sewing functional
that cyclically identifies the open-cut data,
\begin{equation}
    \xi_{A,k}^-
    =
    \xi_{A,k+1}^+,
    \qquad
    k=1,\ldots,n,
    \qquad
    \xi_{A,n+1}^+\equiv\xi_{A,1}^+.
    \label{eq:cyclic-open-cut-sewing}
\end{equation}
It acts on $n$ cut-open boundary functionals and produces the
corresponding replicated boundary amplitude,
\begin{equation}
    \mathfrak S_A^{(n)}
    \left[
        Z_{\lambda_1,A},
        \ldots,
        Z_{\lambda_n,A}
    \right].
    \label{eq:replica-sewing-functional}
\end{equation}
When the boundary data admit a microscopic realization,
$\mathfrak S_A^{(n)}$ reproduces the contraction induced by the cyclic
permutation operator $\mathbb V_A^{(n)}$.

Gravitationally, this sewing is defined most generally on the full
$n$-fold cover. The replica boundary conditions are invariant under
the cyclic group $\mathbb Z_n$, but an individual bulk filling need
not preserve this symmetry. When a saddle does preserve
$\mathbb Z_n$, it may equivalently be described by its quotient. In
Einstein gravity, the fixed locus in the quotient is represented by a
cosmic brane anchored at $\partial A$ with tension
$T_n=(n-1)/(4nG_N)$. Replica-symmetry-breaking saddles must instead be
treated on the full $n$-fold geometry and cannot be represented by a
single cosmic-brane quotient.

The gravitational sewing functional can therefore be defined whenever
the required cut-open wavefunctionals and their gluing data are
available, independently of whether they possess a conventional
boundary-theory interpretation. Away from a consistent microscopic
realization, however, the resulting amplitude need not be
interpretable as the trace of a density matrix. The relation between
the gravitational space of wavefunctions and the subset admitting
consistent CFT interpretations is discussed in
Subsection~\ref{sec:CFT}.

Applying the sewing functional to $n$ boundary values of the cut-open
universe field defines the composite replica operator
\begin{equation}
    \widehat{\mathcal R}_n
    :=
    \mathfrak S_A^{(n)}
    \left[
        \mathcal B_A\widehat\Psi_{A,1},
        \ldots,
        \mathcal B_A\widehat\Psi_{A,n}
    \right].
    \label{eq:third-quantized-replica-operator}
\end{equation}
The replica labels distinguish the copies until the sewing has been
performed. Thus, $\widehat{\mathcal R}_n$ does not permute entire
universes, which would act trivially on the bosonic universe Fock
space. Rather, it cyclically identifies the open-cut boundary data
carried by the $n$ labeled universe-field insertions. When the
resulting bulk saddle preserves $\mathbb Z_n$, one may instead use the
quotient description in terms of a cosmic-brane-decorated
one-universe wavefunction. In that description, the action of
$\mathfrak S_A^{(n)}$ is not applied separately, since its effect is
already encoded in the cosmic-brane monodromy and backreaction.

For continuous boundary fields, the sewing functional takes the
schematic form
\begin{equation}
    \mathfrak S_A^{(n)}
    \left[
        Z_{A,1},\ldots,Z_{A,n}
    \right]
    =
    \int
    \prod_{k=1}^{n}\mathcal D\xi_{A,k}\,
    \prod_{k=1}^{n}
    Z_{A,k}
    \left[
        \xi_{A,k+1},
        \xi_{A,k}
    \right],
    \qquad
    \xi_{A,n+1}\equiv\xi_{A,1}.
    \label{eq:replica-sewing-functional-explicit}
\end{equation}
This is the functional-integral analogue of multiplying $n$ density
matrices and taking their trace. The configurations exposed along the
cuts are integrated over because they become internal sewing data,
whereas the prescribed external boundary data, such as asymptotic
sources or moduli, are held fixed.

These geometries can be organized according to the connectivity of
the $n$ marked replica boundaries:
\begin{equation}
    \mathcal Z_n
    =
    \sum_{\pi\in\operatorname{Part}(n)}
    \mathcal Z_{n,\pi},
    \label{eq:replica-connectivity-decomposition}
\end{equation}
where each block of the partition $\pi$ specifies a collection of
replica boundaries belonging to the same connected bulk component.
The partition into $n$ singleton blocks gives the fully disconnected
contribution $\mathcal Z_n^{\mathrm{disc}}$, while the single-block
partition gives the fully connected replica-wormhole contribution,
\begin{equation}
    \mathcal Z_n^{\mathrm{RW}}
    =
    \langle\Omega|
    U_I^\dagger\,
    \widehat{\mathcal R}_n\,
    U_I
    |\Omega\rangle_{\mathrm{conn}},
    \label{eq:replica-wormhole-correlator}
\end{equation}
with the arguments $(T_{\mathrm f},T_0)$ of $U_I$ left implicit. Here
the connected subscript denotes the cumulant associated with the
single-block connectivity sector.

For integer $n$, the decomposition in 
Eq.~\eqref{eq:replica-connectivity-decomposition} also contains partially
connected sectors corresponding to nontrivial partitions of the
replica labels. There is no general reason to discard them. The
standard two-saddle description of the island transition assumes that
the fully disconnected and fully connected replica-symmetric sectors
dominate in the relevant semiclassical regime
\cite{Penington:2019kki,Almheiri:2019qdq}, so that
\begin{equation}
    \mathcal Z_n
    \simeq
    \mathcal Z_n^{\mathrm{disc}}
    +
    \mathcal Z_n^{\mathrm{RW}}.
    \label{eq:replica-two-saddle-approximation}
\end{equation}
Whether partially connected or replica-symmetry-breaking sectors are
subleading must ultimately be determined from their third-quantized
amplitudes.

The normalized replica moment is
\begin{equation}
    \operatorname{Tr}\rho_A^n
    =
    \frac{\mathcal Z_n}{(\mathcal Z_1)^n},
    \label{eq:normalized-replica-moment}
\end{equation}
and the von Neumann entropy follows from
\begin{equation}
    S_A
    =
    -\left.
    \partial_n
    \log
    \frac{\mathcal Z_n}{(\mathcal Z_1)^n}
    \right|_{n=1}.
    \label{eq:replica-von-neumann-entropy}
\end{equation}
The island transition is governed by the change in dominance between
$\mathcal Z_n^{\mathrm{disc}}$ and
$\mathcal Z_n^{\mathrm{RW}}$. The fully connected contribution cannot
therefore be considered in isolation.

The third-quantized construction provides a common description of the
bulk connectivity sectors entering the replica calculation. The
microscopic permutation $\mathbb V_A^{(n)}$ induces the boundary
sewing functional $\mathfrak S_A^{(n)}$, while the topology-changing
Hamiltonian generates the bulk histories compatible with that sewing.
An explicit treatment of the general case requires the construction
of the cut-open universe field, including its gravitational edge data,
inner product, and sewing measure, as well as the relevant
topology-changing interaction vertices. These ingredients are
substantially more involved than those of the uncut sector considered
below.

The replica-symmetric sector may nevertheless offer a tractable
starting point. When the bulk saddle preserves $\mathbb Z_n$, the
problem reduces to the quantization of a single quotient geometry
containing a cosmic brane. In the AdS$_3$ torus sector, this suggests
an extension of the reduced phase-space construction to punctured
tori with fixed defect holonomy.\footnote{This is an intrinsically
gravitational construction and does not require the corresponding
wavefunctions to admit a consistent CFT interpretation. Such an
interpretation is required only if the resulting replica amplitude is
to be identified microscopically with $\operatorname{Tr}\rho_A^n$; see
Subsection~\ref{sec:CFT}.}

\subsection{Effective Ensemble Interpretation}
\label{sec:ensembles}

The preceding discussion described spacetime-wormhole and
replica-wormhole correlations using radial evolution in the universe
Fock space. 
In that slicing, the wormhole amplitude is encoded in the
pair-correlated component generated by the topology-changing
Hamiltonian. A complementary description is provided by cutting the
same codimension-zero geometry along a reflection-symmetric
Euclidean-time slice. Although this construction was illustrated
explicitly for the torus wormhole in
Subsection~\ref{sec:ER}, its underlying logic is
more general: whenever the cutting surface has two disconnected ER
components, a single connected half-wormhole prepares a state in the
tensor product of two complete one-universe Hilbert spaces. Assuming
that reflection symmetry identifies the corresponding Schmidt bases,
this state takes the TFD-like form
\begin{equation}
    \vert\Psi_{\mathrm{WH},1/2}(\tau)\rangle
    =
    \int d\mu(\lambda)\,
    k_\lambda(\tau)\,
    \vert\lambda\rangle_{\mathrm{ER},1}
    \otimes
    \vert\lambda\rangle_{\mathrm{ER},2}.
    \label{eq:WH-Schmidt-ensemble}
\end{equation}
Here $d\mu(\lambda)$ includes both discrete sums and continuous
integrals, as appropriate. It is this complementary representation of
the wormhole state that makes its effective ensemble interpretation
transparent.

The meaning of the Schmidt label differs from that in an ordinary
thermofield-double state,
\begin{equation}
    \vert\mathrm{TFD}(\beta)\rangle
    =
    \sum_n
    e^{-\beta E_n/2}
    \vert n\rangle_L\otimes\vert n\rangle_R.
\end{equation}
There, $n$ labels a state within a fixed quantum theory, and the two
Hilbert-space factors describe two sectors of a single universe. In
Eq.~\eqref{eq:WH-Schmidt-ensemble}, by contrast, each factor
$\vert\lambda\rangle_{\mathrm{ER}}$ represents a complete ER
geometry, or equivalently a complete one-universe wavefunction. The
label $\lambda$ therefore characterizes an entire one-universe
sector, rather than a state propagating within a fixed universe.

This distinction becomes particularly concrete in the torus example.
The boundary value of a one-universe wavefunction is a candidate torus partition function of the form
\begin{equation}
    Z_\lambda(\tau)
    =
    Z_{\mathrm{light}}(\tau)
    +
    \sum_A
    \alpha_A^{(\lambda)}\psi_A(\tau),
    \label{eq:CFT-partition-lambda}
\end{equation}
where the sum includes the continuous part of the normalizable
spectrum. The complete set of coefficients
$\{\alpha_A^{(\lambda)}\}$ specifies the normalizable, or heavy, part
of the CFT data. It is therefore natural to interpret $\lambda$ as
labeling a candidate CFT realization rather than an individual state within one
CFT.

For each Schmidt sector, let $Z_\lambda(\tau)$ denote the boundary
value of the corresponding complete one-universe wavefunction, where
$\tau$ is allowed to be complex. The coefficient of this sector in the
half-wormhole state may then be written as
\begin{equation}
    k_\lambda(\tau)
    =
    c_\lambda Z_\lambda(\tau),
\end{equation}
where $c_\lambda$ is a $\tau$-independent preparation amplitude.
Gluing the half-geometry to its reflection-related conjugate gives
\begin{align}
    Z_{\mathrm{WH}}(\tau,\bar{\tau})
    =
    \left\langle
    \Psi_{\mathrm{WH},1/2}(\tau)
    \middle|
    \Psi_{\mathrm{WH},1/2}(\tau)
    \right\rangle
    =
    \int_\Lambda d\mu(\lambda)\,
    |c_\lambda|^2\,
    Z_\lambda(\bar{\tau})Z_\lambda(\tau),
    \label{eq:WH-norm-ensemble}
\end{align}
where reflection implies
$Z_\lambda(\bar{\tau})=Z_\lambda(\tau)^*$ on the
reflection-symmetric contour. Absorbing the fixed preparation weights
into the measure,
\begin{equation}
    d\nu(\lambda)
    =
    |c_\lambda|^2d\mu(\lambda),
\end{equation}
we obtain
\begin{equation}
    Z_{\mathrm{WH}}(\tau,\bar{\tau})
    =
    \int_\Lambda d\nu(\lambda)\,
    Z_\lambda(\tau)Z_\lambda(\bar{\tau}).
    \label{eq:WH-ensemble}
\end{equation}
Here $\tau$ denotes the complexified boundary parameter conjugate to
the field-theory energy, rather than necessarily the conventional
upper-half-plane modulus of a Euclidean torus. For the
spectral-form-factor continuation,
\begin{equation}
    \tau=\beta+it,
    \qquad
    \bar{\tau}=\beta-it,
\end{equation}
Eq.~\eqref{eq:WH-ensemble} becomes
\begin{equation}
    Z_{\mathrm{WH}}(\beta,t)
    =
    \int_\Lambda d\nu(\lambda)\,
    Z_\lambda(\beta+it)Z_\lambda(\beta-it).
    \label{eq:WH-ensemble-SFF}
\end{equation}
Thus, the norm of the geometrically prepared two-universe state has
the same form as an ensemble-averaged spectral form factor. Provided
that $\lambda$ labels distinct assignments of complete boundary
theory data, $d\nu(\lambda)$ may be interpreted as an effective
ensemble measure. This interpretation is not imposed as an
independent assumption about the boundary theory; it follows from
resolving the half-wormhole state into sectors labeled by complete
one-universe wavefunctions.

The same structure may be expressed in terms of a reduced state.
Tracing over either ER component gives
\begin{equation}
    \rho_{\mathrm{ER},1}(\tau)
    =
    \frac{1}{\mathcal N(\tau)}
    \int d\mu(\lambda)\,
    |k_\lambda(\tau)|^2
    \vert\lambda\rangle\langle\lambda\vert,
    \qquad
    \mathcal N(\tau)
    =
    \int d\mu(\lambda)\,
    |k_\lambda(\tau)|^2.
    \label{eq:reduced-ER-density-matrix}
\end{equation}
The half-wormhole state therefore provides a purification of the
effective statistical distribution over the one-universe sectors.
Tracing is not required to obtain
Eq.~\eqref{eq:WH-ensemble-SFF}; rather, it makes explicit that the
effective mixed description arises by restricting a pure state in
the doubled universe Hilbert space to one of its ER components.

This interpretation requires the Schmidt basis to be identifiable
across the relevant boundary moduli and to correspond, under the
boundary map, to definite boundary-theory data. These conditions are
not generic properties of an arbitrary entangled state. They are
plausible here because the two Schmidt factors represent entire ER
geometries and because, in the torus sector, the boundary value of each
one-universe wavefunction determines a complete candidate partition
function. The resulting spectral decomposition is intrinsically a
statement about the gravitational one-universe Hilbert space. Its
interpretation as an ensemble of CFTs further requires that the
candidate boundary data define consistent individual theories.
Modular invariance and factorization alone do not guarantee reflection
positivity, spectral integrality, crossing symmetry, or higher-genus
sewing consistency.

The construction is closely analogous to the
Coleman--Marolf--Maxfield description in terms of sectors with
definite boundary-theory data
\cite{Coleman:1988cy,Coleman:1988tj,Marolf:2020xie}, but its origin is
different: here the doubled state and its effective distribution arise
directly from cutting a connected spacetime-wormhole geometry. A
similar qualification applies to the $\alpha$-sectors of the
baby-universe formalism. Their factorization property does not by
itself establish that every set of $\alpha$-eigenvalues is the
generating functional of a consistent CFT. In both constructions, the
sector labels are well defined on the gravitational side, while their
interpretation as labels of boundary theories requires additional CFT
consistency conditions.

The ensemble form obtained above is therefore an effective
representation of particular amplitudes within the third-quantized
theory, not an identification of the underlying theory with a
fundamental statistical ensemble of CFTs. We make this distinction
precise in the next subsection.

\subsection{Implications for AdS/CFT}
\label{sec:CFT}

The construction developed above is formulated intrinsically in terms
of physical gravitational wavefunctions and their topology-changing
interactions. Its interpretation through AdS/CFT requires an
additional step. A boundary CFT should not be identified with an
individual basis mode of the one-universe Hilbert space. Rather, its
partition function, or more generally its generating functional,
supplies asymptotic boundary data that select a complete physical
one-universe wavefunction.

For fixed non-normalizable background data, it is useful to introduce
the affine space of complete one-universe wavefunctions,
\begin{equation}
    \mathfrak A_{\mathrm U}
    :=
    \Psi_{\mathrm{bg}}
    +
    \mathcal H_{\mathrm U},
\end{equation}
where $\mathcal H_{\mathrm U}$ is the normalizable physical
one-universe Hilbert space. Every element of
$\mathfrak A_{\mathrm U}$ represents gravitationally admissible data.
The subset whose boundary values define consistent CFTs forms the
CFT-realizable locus
\begin{equation}
    \mathfrak M_{\mathrm{CFT}}
    :=
    \left\{
        \Psi\in\mathfrak A_{\mathrm U}
        \ \middle|\
        \mathcal B[\Psi]\ \text{defines a consistent CFT}
    \right\}
    \subseteq
    \mathfrak A_{\mathrm U},
    \label{eq:CFT-realizable-locus}
\end{equation}
where $\mathcal B$ denotes the asymptotic boundary map. Thus,
\begin{equation}
    \mathfrak M_{\mathrm{CFT}}
    \subseteq
    \mathfrak A_{\mathrm U}
    =
    \Psi_{\mathrm{bg}}+\mathcal H_{\mathrm U}.
    \label{eq:CFT-gravity-hierarchy}
\end{equation}
The locus $\mathfrak M_{\mathrm{CFT}}$ need not be a linear subspace.
Whether the inclusion in
Eq.~\eqref{eq:CFT-gravity-hierarchy} is strict is a nontrivial question
raised by the third-quantized construction. Generic elements of
$\mathfrak A_{\mathrm U}$ need not satisfy the known conditions for an
exact conventional CFT interpretation, but this does not make them
inconsistent gravitational states or exclude a more general
holographic description.

Operationally, constructing the one-universe field does not require a
classification of the consistent CFTs themselves. Let
$\{\psi_A(q,T)\}$ be a complete orthonormal basis of
$\mathcal H_{\mathrm U}$, and let
\begin{equation}
    \phi_A(q)
    :=
    \mathcal B[\psi_A](q)
\end{equation}
denote the corresponding boundary functions. Equivalently, if the
bulk modes are reconstructed from the boundary, one requires a radial
deformation that extends each boundary basis element to a physical
bulk wavefunction,
\begin{equation}
    \phi_A(q)
    \longmapsto
    \psi_A(q,T),
    \qquad
    \psi_A(q,T_0)
    =
    \phi_A(q).
    \label{eq:boundary-bulk-mode-map}
\end{equation}
The universe field is then constructed directly as
\begin{equation}\label{def_UF}
    \widehat\Psi(q,T)
    =
    \sum_A
    \psi_A(q,T)a_A.
\end{equation}
The basis functions are therefore building blocks from which CFT data
may be composed; they need not themselves be partition functions of
individual CFTs.

A coherent state is specified by
\begin{equation}
    a_A|\Omega_\alpha\rangle
    =
    \alpha_A|\Omega_\alpha\rangle.
\end{equation}
A particular boundary theory is selected by choosing the parameters
$\alpha_A$ such that
\begin{align}
    \Psi_{\mathrm{phys}}(q,T)
    =
    \Psi_{\mathrm{bg}}(q,T)
    +
    \langle\Omega_\alpha|
    \widehat\Psi(q,T)
    |\Omega_\alpha\rangle
    =
    \Psi_{\mathrm{bg}}(q,T)
    +
    \sum_A
    \alpha_A\psi_A(q,T)
    \label{eq:CFT-selected-wavefunction}
\end{align}
has the partition function, or more generally the generating
functional, of the CFT under consideration as its asymptotic boundary
value. Thus, consistent boundary data are defined by the complete
wavefunction selected by the coherent state together with the fixed
non-normalizable contribution, rather than by each gravitational mode
separately. The CFT-realizable locus in
Eq.~\eqref{eq:CFT-realizable-locus} may accordingly be regarded as a
restricted locus in the space of coherent-state parameters.

Against this background, a related possibility is suggested by the
averaged large-$N$ prescription of Kudler-Flam and Witten and the
filtering proposal of
Liu~\cite{Kudler-Flam:2025cki,Kudler-Flam:2026nzz,Liu:2025ikq,Liu:2026fnd}
(see the Introduction for related results of Gesteau and Perlmutter
bearing on the status of this large-$N$
limit~\cite{Gesteau:ClosedUniverses,Perlmutter:BlackHoles}).
Let $\Psi_{\mathrm{CFT}}\in\mathfrak M_{\mathrm{CFT}}$ satisfy
\begin{equation}
    \mathcal B[\Psi_{\mathrm{CFT}}]
    =
    Z_{\mathrm{CFT}}
    =
    \mathbb F[Z_{\mathrm{CFT}}]
    +
    \delta_e Z_{\mathrm{CFT}},
    \qquad
    \mathbb F[\delta_e Z_{\mathrm{CFT}}]=0,
    \label{eq:filtered-boundary-decomposition}
\end{equation}
where $\mathbb F$ may be implemented, for example, by Mellin
averaging. We propose that the smooth part
reconstructs the fixed non-normalizable background,
\begin{equation}
    \mathcal B[\Psi_{\mathrm{bg}}]
    =
    \mathbb F[Z_{\mathrm{CFT}}],
    \label{eq:filtered-background-reconstruction}
\end{equation}
whereas the oscillatory remainder is carried by the quantized
normalizable sector,
\begin{equation}
    \Psi_{\mathrm{CFT}}
    =
    \Psi_{\mathrm{bg}}
    +
    \sum_A\alpha_A^{\mathrm{CFT}}\psi_A,
    \qquad
    \mathcal B\!\left[
        \sum_A\alpha_A^{\mathrm{CFT}}\psi_A
    \right]
    =
    \delta_e Z_{\mathrm{CFT}}.
    \label{eq:erratic-normalizable-identification}
\end{equation}
At this level, the filter therefore supplies a convention for fixing
the affine representative whose ambiguity was discussed around
Eq.~\eqref{eq:background-representative-shift}.
This is a sector-level identification for the complete boundary
observable, not a coefficientwise action of $\mathbb F$.

Filtering then suggests a notion of holographic realizability weaker
than exact CFT realizability. Within the fixed-background sector,
define
\begin{equation}
    \mathfrak M_{\mathbb F}
    :=
    \left\{
        \Psi\in\mathfrak A_{\mathrm U}
        \ \middle|\
        \mathbb F[\mathcal B[\Psi]]
        =
        \mathbb F[\mathcal B[\Psi_{\mathrm{CFT}}]]
        \ \text{for some}\
        \Psi_{\mathrm{CFT}}\in\mathfrak M_{\mathrm{CFT}}
    \right\},
    \label{eq:filtered-holographic-locus}
\end{equation}
where the filter is assumed to extend to the large-$N$ families
defined by the candidate boundary data. By construction,
\begin{equation}
    \mathfrak M_{\mathrm{CFT}}
    \subseteq
    \mathfrak M_{\mathbb F}
    \subseteq
    \mathfrak A_{\mathrm U}.
    \label{eq:filtered-holographic-enlargement}
\end{equation}
The size of this gap --- the erratic content filtered out in passing
from an exact CFT to its filtered image --- may be compared to the
energy-window precision limit of order $S_0 e^{-S_0}$ found by
Perlmutter via extreme-value statistics at fixed $N$, though the
relation between energy-window and large-$N$ erraticness remains to
be established~\cite{Perlmutter:BlackHoles}.
Either inclusion may be strict.
An exact CFT selects a particular set of normalizable coefficients, whereas
$\mathfrak M_{\mathbb F}$ may contain other normalizable completions
with the same filtered boundary image. The full gravitational space
$\mathfrak A_{\mathrm U}$ additionally contains arbitrary
normalizable data whose boundary contributions need not be
filter-null.

For a candidate
$\Psi\in\mathfrak M_{\mathbb F}$ belonging to a large-$N$ family,
write
\begin{equation}
    \Psi
    =
    \Psi_{\mathrm{bg}}
    +
    \delta\Psi,
    \qquad
    \delta\Psi
    =
    \sum_A
    \alpha_A\psi_A.
\end{equation}
The coefficients define a coherent seed and its interaction-picture
evolution,
\begin{equation}
    a_A|\Omega\rangle
    =
    \alpha_A|\Omega\rangle,
    \qquad
    |\Omega(T)\rangle
    =
    U_I(T,T_0)|\Omega\rangle.
    \label{eq:evolved-filtered-coherent-state}
\end{equation}
For the CFT-realizable completion, one has
$\alpha_A=\alpha_A^{\mathrm{CFT}}$,, 
but the filtered locus may contain
other completions. These eigenvalues need not vanish at fixed $N$;
the filter-null condition applies to the complete oscillatory
one-boundary contribution.

In the remainder of this subsection, we denote the normalizable
universe field $\widehat\Psi$ defined in Eq.~\eqref{def_UF} by
$\widehat{\delta\Psi}$. Its expectation value in the unevolved
coherent seed reproduces the normalizable wavefunction,
\begin{equation}
    \langle\Omega|
    \widehat{\delta\Psi}(q,T)
    |\Omega\rangle
    =
    \delta\Psi(q,T).
\end{equation}
After topology-changing evolution, its dressed mean is
\begin{equation}
    \delta\Psi^{\mathrm{dress}}(q,T_f)
    :=
    \langle\Omega|
    U_I^\dagger(T_f,T_0)\,
    \widehat{\delta\Psi}(q,T_f)\,
    U_I(T_f,T_0)
    |\Omega\rangle.
\end{equation}
We center only the normalizable sector,
\begin{equation}
    \widehat{\delta\Psi}_{\mathrm c}(q,T_f)
    :=
    \widehat{\delta\Psi}(q,T_f)
    -
    \delta\Psi^{\mathrm{dress}}(q,T_f),
    \label{eq:centered-universe-field}
\end{equation}
leaving the fixed non-normalizable background unchanged. The complete
centered physical field is therefore
\begin{equation}
    \widehat\Psi_{\mathrm{phys},\mathrm c}(q,T_f)
    :=
    \Psi_{\mathrm{bg}}(q,T_f)
    +
    \widehat{\delta\Psi}_{\mathrm c}(q,T_f).
\end{equation}
By construction, the expectation value of
$\widehat{\delta\Psi}_{\mathrm c}$ in the fully evolved state
vanishes, whereas that of
$\widehat\Psi_{\mathrm{phys},\mathrm c}$ equals
$\Psi_{\mathrm{bg}}$. Thus, at the level of correlators, centering subtracts the complete
dressed normalizable one-universe mean while leaving the fixed
background intact.

Let $Z_i$ denote boundary observables extracted from the same
large-$N$ family, and define
\begin{equation}
    \delta_e Z_i
    :=
    Z_i-\mathbb F[Z_i],
    \qquad
    \mathbb F[\delta_e Z_i]=0,
\end{equation}
Correspondingly, whenever the required radial reconstruction exists,
let $\widehat{\delta\Psi}_{\mathrm c,i}$ denote the bulk insertion
associated with $Z_i$.
We then propose the schematic matching
\begin{equation}
    \mathbb F\!\left[
        \delta_e Z_{i_1}\cdots
        \delta_e Z_{i_n}
    \right]_{\mathrm{conn}}
    \quad\longleftrightarrow\quad
    \mathcal A_{\mathrm U,\mathrm{conn}}^{(n)}
    \!\left[
        \widehat{\delta\Psi}_{\mathrm c,i_1},
        \ldots,
        \widehat{\delta\Psi}_{\mathrm c,i_n}
    \right],
    \qquad
    n\geq1,
    \label{eq:filtered-multiuniverse-correspondence}
\end{equation}
where all insertions on the right-hand side are evaluated in the same
evolved state $U_I(T_f,T_0)|\Omega\rangle$. For $n=1$, both sides
vanish by construction; for $n\geq2$, subtracting the mean leaves
connected amplitudes unchanged. 
Centering therefore subtracts only the dressed one-universe mean from
the correlators; it leaves connected multi-universe amplitudes
unchanged, including their handle and higher-topology contributions.

Membership in $\mathfrak M_{\mathbb F}$ fixes only the filtered
one-boundary image and does not guarantee this higher-point matching.
When satisfied, the correspondence constrains both the coherent-state
eigenvalues specifying the normalizable completion and the splitting
and joining interactions in the third-quantized Hamiltonian. These
constraints need not determine the eigenvalues uniquely, since
distinct completions may yield the same surviving filtered hierarchy.

These observations also allow us to return to the distinction
anticipated at the end of the preceding subsection between a
statistical ensemble of microscopic CFTs and the state space of the
third-quantized theory. Such an ensemble is supported on the exact
CFT-realizable locus $\mathfrak M_{\mathrm{CFT}}$, whereas filtered
holographic realizability may extend to
$\mathfrak M_{\mathbb F}$ and third quantization is formulated on the
full gravitational one-universe space $\mathfrak A_{\mathrm U}$ and
its associated Fock space. The spectral resolution discussed above
therefore defines, in general, an effective ensemble over
gravitational one-universe sectors. It can be interpreted literally
as an ensemble of consistent CFTs only when its support lies in
$\mathfrak M_{\mathrm{CFT}}$; this is an additional condition not
implied by the third-quantized construction. Generic
interaction-generated states may involve wavefunctions outside this
locus, so the effective ensemble interpretation is a property of
particular amplitudes rather than the definition of the underlying
theory.

The torus sector, which underlies the spectral representation
discussed above and will be developed explicitly in
Section~\ref{sec:AdS3}, makes this distinction concrete. Modular
invariance alone does not place a wavefunction on
$\mathfrak M_{\mathrm{CFT}}$: its boundary value must also admit a
positive spectral interpretation, with nonnegative integer
multiplicities for a discrete spectrum, while a complete CFT further
requires compatible higher-point functions, crossing symmetry, OPE
associativity, and higher-genus consistency. The gravitational
construction is well defined throughout $\mathfrak A_{\mathrm U}$,
whereas a conventional CFT interpretation is established only on
$\mathfrak M_{\mathrm{CFT}}$.

Only the normalizable sector is third-quantized in the torus
construction. Its modes modify the spectrum above the black-hole
threshold, while the vacuum and light spectrum remain fixed
non-normalizable data, consistently with the proposal of Schlenker
and Witten that ensemble-like freedom may be confined to the
black-hole sector~\cite{Schlenker:2022dyo}. Because its orthogonal
basis and radial evolution are known explicitly, the torus
one-universe field can be constructed without first classifying the
consistent CFTs assembled from its modes. CFT consistency then selects
distinguished coherent states within this larger gravitational space
and thereby supplies initial data for radial
evolution.\footnote{A related state-selection role is implicit in the
conventional $\alpha$-state formalism. Fixing an $\alpha$-sector
assigns definite eigenvalues $Z_\alpha[J]$ to boundary-creation
operators and may, when these eigenvalues define consistent boundary
data, be regarded as selecting a gravitational wavefunction.}
Whether the remaining wavefunctions admit a more general holographic
interpretation is an open question to which we return in
Section~\ref{sec:discussion}.
\section{A Concrete Example: The Torus Sector of AdS$_3$ Gravity}
\label{sec:AdS3}

We now apply the general third-quantized construction to AdS$_3$
gravity in the torus sector \cite{Moncrief:1989dx,Hosoya:1989yj,Fujiwara:1989xg,
Ezawa:1993ti,Carlip:2004ba,Carlip:1991ij,Carlip:1994ap,
Carlip:1992cj}. This example is particularly tractable:
the physical one-universe Hilbert space admits an explicit modular
spectral decomposition, its radial evolution is known, and the
two-torus spacetime-wormhole amplitude provides concrete information
about topology-changing interactions. It therefore allows us to test
whether a known wormhole amplitude can be embedded consistently into
the third-quantized sewing framework developed above.

We interpret the Cotler--Jensen (CJ) torus wormhole
\cite{Cotler:2020ugk} as the leading, handle-free connected
two-universe contribution generated by a cubic splitting vertex. More precisely, it is obtained from the general mode expansion
\eqref{eq:WH-wavefunction-mode-expansion},
\begin{equation}
    \mathcal W_{\mathrm{CJ}}(1,2)
    =
    \sum_{A,B}
    \psi_A(q_1,T_0)
    \psi_B(q_2,T_0)
    C_{AB}^{\mathrm{CJ}},
    \label{eq:CJ-wavefunction-mode-expansion}
\end{equation}
where $C_{AB}^{\mathrm{CJ}}$ is identified with the leading-order
coefficient in Eq.~\eqref{eq:WH-leading-pair-matrix}. The CJ amplitude
therefore constrains the cubic splitting vertex but does not determine
the complete third-quantized Hamiltonian.

The full connected two-universe matrix also receives higher-order
contributions,
\begin{equation}
    C_{AB}^{\mathrm{WH}}
    =
    C_{AB}^{\mathrm{CJ}}
    +
    C_{AB}^{\mathrm{handle}}
    +
    \cdots ,
    \label{eq:full-WH-pair-matrix-expansion}
\end{equation}
where $C_{AB}^{\mathrm{handle}}$ denotes connected geometries with
internal handles generated by repeated splitting and joining
processes, such as the contribution illustrated in
Figure~\ref{fig:loop}. These terms belong to the complete
two-universe amplitude even when they are subleading in the relevant
topological or semiclassical expansion. Higher elementary vertices
and amplitudes with additional exposed universe boundaries contain
further information not fixed by the CJ result.

As reviewed in Appendix~\ref{sec:CQ_York}, reduced phase-space
quantization of asymptotically Euclidean AdS$_3$ gravity with a torus
conformal boundary yields the Schr\"odinger equation
\eqref{eq:Euclidean-WdW-time-Schrodinger},
\begin{equation}
    i\frac{\partial}{\partial T}\Psi(T,m)
    =
    \widehat h_{\mathrm{mod}}\Psi(T,m),
    \qquad
    \widehat h_{\mathrm{mod}}
    :=
    \sqrt{
        -\left(
            \Delta_{\mathrm{Maass}}+\frac14
        \right)
    },
    \label{eq:Schrodinger}
\end{equation}
where
\begin{equation}
    \Delta_{\mathrm{Maass}}
    :=
    m_2^2
    \left(
        \partial_{m_1}^2+\partial_{m_2}^2
    \right),
\end{equation}
and we have adopted the operator ordering used in
Ref.~\cite{Hirano:2026lpp}. The shifted Maass Laplacian admits a
spectral decomposition in terms of weight-zero automorphic
eigenfunctions,
\begin{equation}
    \left(
        \Delta_{\mathrm{Maass}}+\frac14
    \right)
    \phi_\lambda(m)
    =
    -r_\lambda^2\phi_\lambda(m),
    \qquad
    \widehat h_{\mathrm{mod}}\phi_\lambda(m)
    =
    r_\lambda\phi_\lambda(m),
    \label{eq:Maass-eigenvalue-equation}
\end{equation}
where $\lambda$ collectively labels both the continuous Eisenstein
sector and the discrete sector of Maass cusp forms. Accordingly, a
normalizable one-universe wavefunction may be expanded as
\begin{align}
    \Psi_{\mathrm{norm}}(T,m)
    =
    \int_0^\infty d\mu_{\mathrm E}(r)\,
    \alpha_{\mathrm E}(r)\,
    e^{-irT}\,
    \phi_r^{(\mathrm E)}(m)
    +
    \sum_n
    \alpha_n\,
    e^{-ir_nT}\,
    \phi_n^{(\mathrm{cusp})}(m).
    \label{eq:torus-one-universe-expansion}
\end{align}
The measure $d\mu_{\mathrm E}(r)$ depends on the normalization chosen
for the Eisenstein series. The coefficients
$\alpha_{\mathrm E}(r)$ and $\alpha_n$ specify the normalizable part
of the one-universe wavefunction.
The fixed non-normalizable contribution
$\Psi_{\mathrm{bg}}(T,m)$ is also modular invariant. Unlike
$\Psi_{\mathrm{norm}}$, however, it need not lie in the function space
to which the Roelcke--Selberg spectral decomposition applies and
therefore need not admit an expansion in the Eisenstein and Maass
cusp-form basis displayed above. Thus, modular invariance alone does
not imply membership in the normalizable one-universe Hilbert space.
In the AdS$_3$ torus example, this distinction is naturally associated
with the separation between the vacuum and light spectrum below the
BTZ black-hole threshold and the heavy spectrum above it
\cite{Maloney:2007ud}. We treat the former as fixed
non-normalizable background data, while the normalizable spectral
sector describes dynamical modifications of the heavy spectrum.

From the boundary-CFT perspective, this decomposition is closely
related to the harmonic decomposition of torus partition functions
developed in Ref.~\cite{Benjamin:2021ygh} and subsequently applied to
wormhole spectral correlations in
Refs.~\cite{Haehl:2023tkr,Haehl:2023xys,DiUbaldo:2023qli}. After
stripping off the universal Virasoro-descendant contribution, the
primary-counting partition function may be written schematically as
\begin{equation}
    Z_{\mathrm P}
    =
    Z_{\mathrm{light}}^{\mathrm{mod}}
    +
    Z_{\mathrm{spec}},
    \label{eq:CFT-light-spectral-decomposition}
\end{equation}
where $Z_{\mathrm{light}}^{\mathrm{mod}}$ is a modular completion
determined by the light spectrum and $Z_{\mathrm{spec}}$ is the
remaining square-integrable spectral contribution.\footnote{The primary-counting partition function should not be
identified with the Maloney--Witten Poincar\'e sum
\cite{Maloney:2007ud}. The former is obtained from the partition
function of a CFT by removing the universal Virasoro-descendant
factor. The latter is a gravitational modular completion of the
vacuum contribution and, after the same descendant factor is removed,
is naturally associated with
$Z_{\mathrm{light}}^{\mathrm{mod}}$ in the vacuum-only pure-gravity
case. It need not equal the complete primary-counting partition
function, which may contain an additional square-integrable
contribution $Z_{\mathrm{spec}}$.}
In the gravitational description, we identify
\begin{equation}
    \Psi_{\mathrm{bg}}
    \longleftrightarrow
    Z_{\mathrm{light}}^{\mathrm{mod}},
    \qquad
    \Psi_{\mathrm{norm}}
    \longleftrightarrow
    Z_{\mathrm{spec}}.
    \label{eq:gravity-CFT-background-spectral-map}
\end{equation}
Thus, the coefficients $\alpha_{\mathrm E}(r)$ and $\alpha_n$ are the
modular spectral coefficients of $Z_{\mathrm{spec}}$, whereas the
light-determined modular completion remains fixed and is not promoted
to a universe field operator. 
In the torus sector, this correspondence can be made explicit using
the integral transform derived in Ref.~\cite{Hirano:2026lpp}
(see also Ref.~\cite{Coleman:2020jte} for an earlier,
closely related construction), which
maps the $T\bar T$-deformed torus partition function to the bulk
one-universe wavefunction.\footnote{The radial reconstruction in
Ref.~\cite{Hirano:2026lpp} was written in terms of
$\cos(rT)$, corresponding to the real combination of the two
Wheeler--DeWitt frequency branches. The reduced Schr\"odinger
evolution used here selects the mode $e^{-irT}$. The corresponding
reconstruction is obtained by supplementing the cosine transform with
its sine transform, or equivalently by applying the appropriate
Hilbert-transform frequency projection.} 
Denoting this radial transform
schematically by $\mathcal K_T$, the two sectors are related by
\begin{equation}
    \Psi_{\mathrm{bg}}(T,m)
    =
    \mathcal K_T
    \!\left[
        Z_{\mathrm{light}}^{\mathrm{mod}}
    \right](m),
    \qquad
    \Psi_{\mathrm{norm}}(T,m)
    =
    \mathcal K_T
    \!\left[
        Z_{\mathrm{spec}}
    \right](m),
    \label{eq:gravity-CFT-background-spectral-map}
\end{equation}
where \begin{equation}
    \mathcal K_T
    \!\left[
        \phi_\lambda(m)
    \right]
    =
    e^{-ir_\lambda T}\phi_\lambda(m).
    \label{eq:radial-transform-of-modular-mode}
\end{equation}
At the asymptotic boundary, this transform reduces to the
corresponding boundary partition-function data. Thus, the coefficients
$\alpha_{\mathrm E}(r)$ and $\alpha_n$ are equivalently the modular
spectral coefficients of $Z_{\mathrm{spec}}$ and the expansion
coefficients of its radially evolved bulk wavefunction. The
light-determined modular completion is mapped to the fixed
non-normalizable bulk contribution and is not promoted to a universe
field operator.

Following the general construction of
Section~\ref{sec:Wormhole3rdQ}, only the normalizable part is promoted
to a universe field operator. In the torus sector, its mode expansion
is
\begin{align}
    \widehat\Psi(T,m)
    =
    \int_0^\infty d\mu_{\mathrm E}(r)\,
    e^{-irT}\,
    \phi_r^{(\mathrm E)}(m)\,
    a_{\mathrm E}(r)
    +
    \sum_n
    e^{-ir_nT}\,
    \phi_n^{(\mathrm{cusp})}(m)\,
    a_n,
    \label{eq:third-quantized-torus-field}
\end{align}
where
\begin{equation}
    \bigl[
        a_{\mathrm E}(r),
        a_{\mathrm E}^\dagger(r')
    \bigr]
    =
    \delta_{\mu_{\mathrm E}}(r,r'),
    \qquad
    [a_n,a_{n'}^\dagger]
    =
    \delta_{nn'},
\end{equation}
and $\delta_{\mu_{\mathrm E}}(r,r')$ is normalized with respect to
$d\mu_{\mathrm E}(r)$.

The corresponding one-universe Hilbert space is the normalizable
modular spectral space
\begin{equation}
    \mathcal H_{\mathrm U}^{(T^2)}
    =
    \int_0^\infty{}^\oplus
    d\mu_{\mathrm E}(r)\,
    \mathcal H_r^{(\mathrm E)}
    \oplus
    \bigoplus_n
    \mathcal H_n^{(\mathrm{cusp})}
    =
    \mathcal H_{\mathrm E}
    \oplus
    \mathcal H_{\mathrm{cusp}}.
    \label{eq:torus-one-universe-Hilbert-space}
\end{equation}
Here $\mathcal H_{\mathrm E}$ denotes the direct integral over the
continuous Eisenstein spectrum, while
$\mathcal H_{\mathrm{cusp}}$ denotes the direct sum over the discrete
Maass cusp-form spectrum.
The fixed modular-invariant contribution
$\Psi_{\mathrm{bg}}$ does not belong to
$\mathcal H_{\mathrm U}^{(T^2)}$ and therefore carries no creation or
annihilation operators. For a coherent state
$\vert\Omega_\alpha\rangle$, the eigenvalues
$\alpha_{\mathrm E}(r)$ and $\alpha_n$ reproduce the coefficients in
Eq.~\eqref{eq:torus-one-universe-expansion}, so that the complete
one-universe wavefunction is recovered by adding
$\Psi_{\mathrm{bg}}$ to
$\langle\Omega_\alpha|\widehat\Psi|\Omega_\alpha\rangle$.

The free one-universe modes
$\psi_A(q,T)$ appearing in
Eq.~\eqref{eq:CJ-wavefunction-mode-expansion} are therefore identified
as
\begin{equation}
    \psi_A(q,T)
    =
    \begin{cases}
        e^{-irT}\,\phi_r^{(\mathrm E)}(m),
        & A=({\mathrm E},r), \\[2mm]
        e^{-ir_nT}\,\phi_n^{(\mathrm{cusp})}(m),
        & A=({\mathrm{cusp}},n),
    \end{cases}
    \qquad\mbox{and}\qquad
    q=m.
    \label{eq:torus-mode-identification}
\end{equation}
Accordingly, the collective sum over one-universe modes becomes
\begin{equation}
    \sum_A
    =
    \int_0^\infty d\mu_{\mathrm E}(r)
    +
    \sum_n.
    \label{eq:torus-spectral-sum}
\end{equation}
With this dictionary established, we now show that the diagonal
modular spectral representation of the CJ wormhole is compatible with
a splitting interaction satisfying the sewing condition.

\subsection{A Sewing-Consistent Diagonal Vertex}
\label{sec:diagonal-torus-vertex}

The CJ amplitude is diagonal in the modular spectral basis
\cite{Haehl:2023tkr,Haehl:2023xys,DiUbaldo:2023qli}. Let
$\lambda$ collectively denote the continuous Eisenstein and discrete
Maass labels, normalized by
\begin{equation}
    \langle\lambda\vert\lambda'\rangle
    =
    \delta_\mu(\lambda,\lambda'),
    \qquad
    \int d\mu(\lambda')\,
    \delta_\mu(\lambda,\lambda')\,
    f(\lambda')
    =
    f(\lambda).
    \label{eq:modular-spectral-delta}
\end{equation}
Thus, $\delta_\mu$ reduces to a Dirac delta in the continuous sector
and a Kronecker delta in the discrete sector. In this basis, the
leading two-universe coefficient matrix takes the form
\begin{equation}
    C_{\lambda\lambda'}^{\mathrm{CJ}}
    =
    \kappa_\lambda^{\mathrm{CJ}}\,
    \delta_\mu(\lambda,\lambda'),
    \label{eq:CJ-diagonal-spectral-matrix}
\end{equation}
where, up to the overall normalization convention for the wormhole
amplitude,
\begin{equation}
    \kappa_\lambda^{\mathrm{CJ}}
    =
    \frac{1}{
        2\cosh(\pi r_\lambda)
    }.
    \label{eq:CJ-spectral-weight}
\end{equation}
This expression applies uniformly to the Eisenstein and Maass
cusp-form sectors when the modes are orthonormal. If unnormalized cusp
forms are used, the corresponding coefficient also contains the
inverse Petersson norm.

It is useful to clarify what information this amplitude contains. Let
$\vert m\rangle$ denote the generalized position state associated
with a boundary torus of modulus $m$. In terms of the modular spectral
basis,
\begin{equation}
    \vert m\rangle
    =
    \int d\mu(\lambda)\,
    \phi_\lambda^*(m)\,
    \vert\lambda\rangle,
    \qquad
    \langle m\vert\lambda\rangle
    =
    \phi_\lambda(m),
    \label{eq:modulus-position-state}
\end{equation}
where the collective integral includes both the continuous and
discrete sectors. After identifying one boundary as incoming by
reflection, the CJ amplitude is the position-space kernel of the
operator
\begin{equation}
    \widehat{\mathcal C}_{\mathrm{CJ}}
    :=
    \int d\mu(\lambda)\,
    \kappa_\lambda^{\mathrm{CJ}}\,
    \vert\lambda\rangle
    \langle\lambda\vert,
    \label{eq:CJ-kernel-operator}
\end{equation}
namely
\begin{equation}
    \mathcal W_{\mathrm{CJ}}(m_1,m_2)
    =
    \langle m_1\vert
    \widehat{\mathcal C}_{\mathrm{CJ}}
    \vert m_2\rangle.
    \label{eq:CJ-position-space-kernel}
\end{equation}
With both boundaries assigned outgoing orientation, the same kernel
is equivalently represented, by vectorization, as a state in the
two-universe Hilbert space. The two descriptions differ only by
orientation and complex-conjugation conventions.

The modulus $m$ specifies a boundary geometry, not a boundary CFT. A
candidate CFT is instead associated with a complete wavefunction,
whose normalizable part has the spectral expansion
\begin{equation}
    \vert Z_{\mathcal C}\rangle_{\mathrm{norm}}
    =
    \int d\mu(\lambda)\,
    \alpha_\lambda^{(\mathcal C)}
    \vert\lambda\rangle.
    \label{eq:CFT-spectral-vector}
\end{equation}
Its complete partition function is recovered only after restoring the
fixed non-normalizable contribution. The CJ kernel itself contains no
coefficients $\alpha_\lambda^{(\mathcal C)}$ selecting a particular
CFT.\footnote{The nonuniqueness discussed in
Refs.~\cite{Haehl:2023tkr,Haehl:2023xys} may have a simple
state-dependent interpretation in the present framework. The CJ
amplitude is the matrix element of the wormhole operator between the
fixed-modulus states $\vert m_1\rangle$ and $\vert m_2\rangle$.
Evaluating the same operator between more general one-universe states
$\vert\Psi_1(m_1)\rangle$ and $\vert\Psi_2(m_2)\rangle$ produces
state-dependent corrections to the CJ result. Whether all the
allowed nonuniversal corrections can be understood in this way
remains to be determined.}
Such data enter only when the kernel is contracted with complete
one-universe wavefunctions. Whether these wavefunctions belong to
$\mathfrak M_{\mathrm{CFT}}$, only to the candidate one-boundary locus
$\mathfrak M_{\mathbb F}$, or lie outside the filtered locus is an
additional question.
The CJ path integral should therefore
be regarded as a gravitational operator on the one-universe Hilbert
space, rather than as a prescription that by itself selects boundary
theories on its two sides.

The diagonal structure in
Eq.~\eqref{eq:CJ-diagonal-spectral-matrix} was obtained through
complementary modular analyses. Refs.~\cite{Haehl:2023tkr,
Haehl:2023xys} derive it by imposing random-matrix spectral
correlations and completing them using number-theoretic trace
formulae, while Ref.~\cite{DiUbaldo:2023qli} formulates the diagonal
pairing as a Hecke projection and constructs a formal spectral square
root of the wormhole amplitude. Here we use this result as
gravitational input and ask whether the corresponding two-leg kernel
admits a factorization through a cubic splitting vertex satisfying the
local sewing condition
\eqref{eq:local-third-quantized-crossing}.

Consider the diagonal splitting ansatz
\begin{equation}
    g_{\lambda_1\lambda_2}{}^{\lambda_3}(T)
    =
    \gamma_{\lambda_1}\,
    \delta_\mu(\lambda_1,\lambda_2)\,
    \delta_\mu(\lambda_1,\lambda_3)\,
    \chi_{\lambda_1}(T).
    \label{eq:diagonal-splitting-ansatz}
\end{equation}
The interaction-picture phase is fixed by the free radial evolution.
For the splitting operator
$a_{\lambda_1}^\dagger
a_{\lambda_2}^\dagger a_{\lambda_3}$, it is
$e^{i(r_{\lambda_1}+r_{\lambda_2}-r_{\lambda_3})T}$.
On the support of the diagonal delta functions, this reduces to
\begin{equation}
    \chi_\lambda(T)
    =
    e^{ir_\lambda T}.
    \label{eq:diagonal-interaction-phase}
\end{equation}
A reversal of the incoming--outgoing convention changes the sign of
the phase without affecting the sewing argument.

After suppressing this known interaction-picture phase, the cubic
tensor defines the reduced product
\begin{equation}
    \vert\lambda_1\rangle
    \star
    \vert\lambda_2\rangle
    :=
    \int d\mu(\lambda_3)\,
    \gamma_{\lambda_1}\,
    \delta_\mu(\lambda_1,\lambda_2)\,
    \delta_\mu(\lambda_1,\lambda_3)\,
    \vert\lambda_3\rangle
    =
    \gamma_{\lambda_1}\,
    \delta_\mu(\lambda_1,\lambda_2)\,
    \vert\lambda_1\rangle.
    \label{eq:diagonal-universe-product}
\end{equation}
This operation describes the joining orientation of the cubic
cobordism and should not be confused with a tensor product. It is
commutative and associative in the distributional sense: either
composition of three modes, $(\vert\lambda_1\rangle\star\vert\lambda_2\rangle)\star\vert\lambda_3\rangle$ or $\vert\lambda_1\rangle\star(\vert\lambda_2\rangle\star\vert\lambda_3\rangle)$, vanishes unless all three labels coincide,
and both give
$\gamma_\lambda^2\vert\lambda\rangle$ in the diagonal sector.

The same structure satisfies the local third-quantized sewing condition~\eqref{eq:local-third-quantized-crossing}. For $0\leq T_1\leq T_2<\infty$,
\begin{equation}
    \int d\mu(E)\,
    g_{CD}{}^E(T_2)\,
    g^{AB}{}_{E}(T_1)
    =
    \int d\mu(E)\,
    g_{BD}{}^E(T_2)\,
    g^{AC}{}_{E}(T_1).
    \label{eq:diagonal-cubic-sewing}
\end{equation}
Both sides vanish unless $A=B=C=D$. When the four labels coincide,
the coefficients and interaction-picture phases agree. The equality
therefore holds pointwise in $T_1$ and $T_2$, and hence also after the
ordered radial integrations. No separate propagator appears because
the free propagation of the intermediate universe is already encoded
in the time dependence of the interaction-picture vertices.

To reproduce the two-leg CJ kernel from this cubic interaction, an
incoming universe must be attached to the third leg and capped by a
specified state. Let
$\vert\Omega_{\mathrm{in}}\rangle$ denote the coherent state prepared
by this cap,
\begin{equation}
    a_\lambda
    \vert\Omega_{\mathrm{in}}\rangle
    =
    \alpha_\lambda^{\mathrm{in}}
    \vert\Omega_{\mathrm{in}}\rangle.
    \label{eq:incoming-cap-coherent-state}
\end{equation}
The coefficients $\alpha_\lambda^{\mathrm{in}}$ characterize the
incoming cap and are independent of any boundary data subsequently
assigned to the two exposed outgoing universes. Contracting the
incoming leg with these coefficients gives the leading connected pair
matrix
\begin{align}
    C_{\lambda\lambda'}^{\mathrm{WH}}
    ={}&
    -i
    \int_0^\infty dT
    \int d\mu(\sigma)\,
    g_{\lambda\lambda'}{}^\sigma(T)\,
    \alpha_\sigma^{\mathrm{in}}
    +
    O\!\left(
        H_{\mathrm{int}}^2
    \right)\nonumber\\
    ={}&
    -i\gamma_\lambda
    \alpha_\lambda^{\mathrm{in}}
    I_\lambda\,
    \delta_\mu(\lambda,\lambda')
    +
    O\!\left(
        H_{\mathrm{int}}^2
    \right),
    \label{eq:diagonal-pair-from-splitting}
\end{align}
where
\begin{equation}
    I_\lambda
    :=
    \int_0^\infty dT\,
    \chi_\lambda(T).
    \label{eq:diagonal-radial-integral}
\end{equation}
For $\chi_\lambda(T)=e^{ir_\lambda T}$, the oscillatory integral is
defined with the usual convergence prescription,
\begin{equation}
    I_\lambda
    :=
    \lim_{\epsilon\to0^+}
    \int_0^\infty dT\,
    e^{ir_\lambda T-\epsilon T}
    =
    \lim_{\epsilon\to0^+}
    \frac{1}{\epsilon-ir_\lambda}.
    \label{eq:regulated-radial-integral}
\end{equation}
For $r_\lambda\neq0$, this gives
\begin{equation}
    I_\lambda
    =
    \frac{i}{r_\lambda}.
\end{equation}

Matching Eq.~\eqref{eq:diagonal-pair-from-splitting} to
Eq.~\eqref{eq:CJ-diagonal-spectral-matrix} determines the contracted
pair source
\begin{equation}
    \Gamma_\lambda^{\mathrm{CJ}}
    :=
    \gamma_\lambda
    \alpha_\lambda^{\mathrm{in}}
    =
    \frac{
        i\kappa_\lambda^{\mathrm{CJ}}
    }{
        I_\lambda
    }
    =
    r_\lambda
    \kappa_\lambda^{\mathrm{CJ}},
    \qquad
    r_\lambda\neq0.
    \label{eq:CJ-contracted-pair-source}
\end{equation}
Using Eq.~\eqref{eq:CJ-spectral-weight}, this becomes
\begin{equation}
    \Gamma_\lambda^{\mathrm{CJ}}
    =
    \frac{
        r_\lambda
    }{
        2\cosh(\pi r_\lambda)
    },
    \label{eq:explicit-CJ-contracted-pair-source}
\end{equation}
up to the overall normalization convention for the CJ amplitude.

The two-boundary CJ path integral fixes
$\Gamma_\lambda^{\mathrm{CJ}}$, but does not separately determine the
uncapped cubic coupling $\gamma_\lambda$ and the state prepared by the
incoming cap. 
In a universal third-quantized theory,
$\gamma_\lambda$ should be regarded as a state-independent coupling
fixed by the gravitational dynamics. Once it has been determined
independently, for example from an appropriate three-boundary
amplitude, matching to the CJ kernel fixes the coherent-state
coefficients of the incoming cap:
\begin{equation}
    \alpha_\lambda^{\mathrm{in}}
    =
    \frac{
        \Gamma_\lambda^{\mathrm{CJ}}
    }{
        \gamma_\lambda
    }
    =
    \frac{
        r_\lambda
        \kappa_\lambda^{\mathrm{CJ}}
    }{
        \gamma_\lambda
    }
    =
    \frac{
        r_\lambda
    }{
        2\gamma_\lambda
        \cosh(\pi r_\lambda)
    },
    \qquad
    \gamma_\lambda\neq0.
    \label{eq:CJ-incoming-cap-coefficient}
\end{equation}
Thus, the CJ amplitude constrains the state prepared by the cap
relative to the universal cubic vertex. If
$\Gamma_\lambda^{\mathrm{CJ}}\neq0$, compatibility requires
$\gamma_\lambda\neq0$ in that spectral sector. Any zero-frequency
sector must instead be treated using
Eq.~\eqref{eq:regulated-radial-integral}.

The incoming cap is not associated with either of the boundary
theories that may subsequently be assigned to the two exposed
outgoing universes. Its coefficients describe the state prepared by
the gravitational cap once the underlying splitting vertex is known.
Whether this state admits an AdS/CFT interpretation is a separate
question and is not required for the gravitational construction. In
particular, for pure AdS$_3$ gravity no microscopic dual CFT is
presently known, and the formal spectral factorization of the
wormhole \cite{DiUbaldo:2023qli} does not by itself establish that the
incoming cap is CFT realizable.

Absent an independent determination of $\gamma_\lambda$, the CJ
amplitude fixes only the contracted source
$\Gamma_\lambda^{\mathrm{CJ}}
=\gamma_\lambda\alpha_\lambda^{\mathrm{in}}$ and therefore admits a
family of factorizations into an uncapped vertex and a cap state. A
universal determination of $\gamma_\lambda$ would require additional
gravitational input, such as an appropriate three-boundary amplitude;
once it is known, the CJ kernel instead determines
$\alpha_\lambda^{\mathrm{in}}$. Higher-boundary amplitudes, loop
corrections, and the complete sewing of cobordism moduli space may
further require additional elementary vertices or non-diagonal
couplings. A minimal completion generated entirely by repeated
splitting and joining remains possible, but is not implied by the
two-boundary result.

The torus sector thus realizes the principal ingredients of the
third-quantized proposal: a physical one-universe Hilbert space, its
free radial evolution, a third-quantized field operator, and a
topology-changing interaction constrained by a known wormhole
amplitude. 
The CJ result fixes the effective two-leg source produced after the
incoming universe is capped and, when extended to a finite radial
endpoint, can supply two-universe input to the third-quantized
Schwinger--Dyson hierarchy.
Determining the
underlying universal interaction and its extension beyond the leading
two-universe sector remains a problem for the complete
third-quantized bootstrap.
 
\section{Comments on Higher-Dimensional Generalizations}
\label{sec:higherD}

The torus example exhibits the principal ingredients required for
third quantization: a physical one-universe Hilbert space, an
orthogonal basis of wavefunctions, a radial evolution law, and
interaction vertices that change bulk connectivity. In higher
dimensions, these ingredients are more difficult to identify because
gravity has local propagating degrees of freedom and the reduced
one-universe Hilbert space is not known explicitly. A tractable
starting point can nevertheless be obtained by restricting attention
to a sector for which an orthogonal harmonic decomposition is already
available.

A complete higher-dimensional one-universe wavefunction depends on
both the boundary insertion data and the physical gravitational
variables $q$, whose boundary description is encoded in the
stress-tensor sector. Resolving this dependence requires an orthogonal
basis for the graviton sector of the reduced gravitational Hilbert
space, corresponding holographically to single- and
multi-stress-tensor states. Linearized gravitons provide perturbative
coordinates on this space around AdS, but a wavefunctional basis
adapted to the reduced radial Hamiltonian is not known in general. As
a first approximation, we fix one component of this sector and write
schematically
\begin{equation}\label{freeze_g}
    \Psi^{(4)}[q;z,\bar z;T]
    \simeq
    \chi_{\mathrm{ref}}(q,T)\,
    G(z,\bar z;T),
\end{equation}
where $\chi_{\mathrm{ref}}(q,T)$ is a chosen reference wavefunction in
the physical graviton sector and $G(z,\bar z;T)$ is the remaining
four-point component, depending on the conformal cross-ratios and
radial time. This truncation freezes transitions to gravitational
wavefunctions orthogonal to $\chi_{\mathrm{ref}}$ but does not imply
that correlators with fewer insertions define trivial bulk
wavefunctions.

At the asymptotic boundary $T=T_0$, we denote the four-point
component introduced above by
$\mathcal G(z,\bar z):=\mathcal G(z,\bar z;T_0)$. Within the
fixed-graviton truncation, this four-point function provides the
simplest nontrivial setting. It may be regarded as a particular
boundary value of the bulk wavefunction, while its conformal
partial-wave decomposition supplies a natural spectral basis,
\begin{equation}
    \mathcal G(z,\bar z)
    =
    \sum_J
    \int_{\mathcal C}d\Delta\,
    c(\Delta,J)\,
    \Phi_{\Delta,J}(z,\bar z),
    \label{eq:CPW-decomposition}
\end{equation}
where $\mathcal C$ denotes the appropriate principal-series contour.
The conformal partial waves are eigenfunctions of the quadratic
Casimir and obey completeness and orthogonality relations on the space
of cross-ratios. They thus provide a higher-dimensional counterpart of
the orthogonal modular basis used in the torus construction. This
four-point sector does not yet constitute a third quantization of the
complete higher-dimensional gravitational theory. It instead provides
a concrete setting in which to investigate the radial deformation of
boundary wavefunctions, the constraints of crossing symmetry, and the
appearance of additional bulk degrees of freedom.

Motivated by the torus example, we seek a radial deformation of this
conformal partial-wave system. We do not assume that the deformation
has been defined microscopically by a local composite $T^2$ operator.
Rather, we use ``$T^2$-type'' to denote a radial deformation whose form
is to be determined by bootstrap. Its defining input consists of the
undeformed conformal partial waves, crossing covariance, closure of
the four-point sector under the radial flow, and the preservation of a
complete orthogonal spectral decomposition. A local field-theoretic
realization, if one exists, would be an outcome rather than an input of
the construction.

The proposed deformation has a coupling $\mu$ of length dimension
$d$,
\begin{equation}
    [\mu]
    =
    (\mathrm{length})^d.
\end{equation}
Because the cross-ratios contain no information about the overall size
of the four-point configuration, the introduction of $\mu$ requires
an additional length scale. A natural choice is supplied by radial
quantization on
\begin{equation}
    \mathbb R\times S_R^{d-1},
\end{equation}
where the sphere radius $R$ plays the role of the torus radius in the
three-dimensional construction. The corresponding dimensionless
coupling is
\begin{equation}
    \hat{\mu}
    =
    \frac{\mu}{R^d}.
    \label{eq:higher-dimensional-dimensionless-coupling}
\end{equation}
Although $R$ may be scaled away in the undeformed CFT, it becomes a
physical scale after the deformation. We use $\hat{\mu}$ to
parametrize the radial continuation away from the asymptotic boundary,
which corresponds to $\hat{\mu}=0$.

To reconstruct bulk wavefunctions from conformal partial waves, we
make the minimal closure ansatz
\begin{equation}
    \Phi_{\Delta,J}(z,\bar z)
    \longmapsto
    \Phi_{\Delta,J}^{(\hat{\mu})}(z,\bar z).
    \label{eq:deformed-CPW-map}
\end{equation}
Namely, we assume that the radial flow defines a distinguished sector
depending only on the cross-ratios and the dimensionless coupling
$\hat{\mu}$. This assumption is necessary for the conformal
partial-wave basis to evolve into a corresponding basis of bulk
wavefunctions. It may be viewed as a spurionic conformal covariance of
the family of deformed theories and, in an elliptic parametrization,
may be implemented by assigning $\hat{\mu}$ an appropriate modular
weight.\footnote{A possible realization is to uniformize the
cross-ratio by $z=\lambda_{\mathrm{mod}}(\tau)$, equivalently
$\tau=iK(1-z)/K(z)$. Under
$\gamma:\tau\mapsto
(a_\gamma\tau+b_\gamma)/(c_\gamma\tau+d_\gamma)$, assigning
$\hat{\mu}\mapsto
\hat{\mu}/|c_\gamma\tau+d_\gamma|^d$
makes
$\xi=(\operatorname{Im}\tau)^{d/2}/\hat{\mu}$ invariant. The
deformation may then be formulated modular covariantly in
$(\tau,\hat{\mu})$, or equivalently crossing covariantly in the
original cross-ratios.}

Within this closed sector, the deformed conformal partial waves should
satisfy
\begin{equation}
    \lim_{\hat{\mu}\to0}
    \Phi_{\Delta,J}^{(\hat{\mu})}(z,\bar z)
    =
    \Phi_{\Delta,J}(z,\bar z),
\end{equation}
transform consistently under crossing, and furnish a complete
orthogonal basis, possibly with a $\hat{\mu}$-dependent inner product.
We further require a diffusion-type flow,
\begin{equation}
    \partial_{\hat{\mu}}
    \Phi_{\Delta,J}^{(\hat{\mu})}
    =
    \mathcal D_{\hat{\mu}}\,
    \Phi_{\Delta,J}^{(\hat{\mu})},
    \label{eq:deformed-CPW-flow}
\end{equation}
where $\mathcal D_{\hat{\mu}}$ is a crossing-covariant differential
operator on the cross-ratios with coefficients that may depend on
$\hat{\mu}$. Equivalently, the deformed waves may be characterized as
eigenfunctions of a deformed Casimir operator,
\begin{equation}
    \mathcal C_{\hat{\mu}}
    \Phi_{\Delta,J}^{(\hat{\mu})}
    =
    c_{\Delta,J}^{(\hat{\mu})}
    \Phi_{\Delta,J}^{(\hat{\mu})}.
    \label{eq:deformed-Casimir-equation}
\end{equation}
The flow operator, the deformed Casimir, and the corresponding inner
product are to be determined jointly from the bootstrap conditions.

Unlike in the torus construction, it is not known whether the
higher-dimensional deformation is unique. 
Within the fixed-graviton sector, the bootstrap conditions may
uniquely determine the flow operator, or they may leave residual
parameters corresponding to bulk couplings, boundary conditions, or
other dynamical data not fixed by crossing and spectral consistency
alone.
Perturbatively, one may write
\begin{equation}
\mathcal C_{\hat{\mu}}
=
\mathcal C_0
+
\hat{\mu}
\sum_I g_I\mathcal O_I
+
O(\hat{\mu}^2),
\label{eq:deformed-Casimir-expansion}
\end{equation}
where the consistency conditions restrict, and may possibly fix, the
operators $\mathcal O_I$ and coefficients $g_I$. This question is
distinct from the closure of the fixed-graviton truncation. If radial
evolution mixes $\chi_{\mathrm{ref}}$ with gravitational wavefunctions
orthogonal to it, the ansatz must instead be enlarged to
\begin{equation}
\Psi^{(4)}[q;z,\bar z;T]
=
\sum_I
\chi_I(q,T),
\mathcal G_I(z,\bar z;T),
\label{eq:unfrozen-graviton-expansion}
\end{equation}
with a flow operator or deformed Casimir that is matrix-valued in
$I$. This enlargement, rather than non-uniqueness within a closed
sector, constitutes the unfreezing of the propagating graviton sector.

If such a deformation can be constructed, the resulting functions provide candidate reduced one-universe modes for the four-point component within the fixed-graviton truncation. 
Parametrizing the principal series by
\begin{equation}
    \Delta
    =
    \frac d2+i\nu,
\end{equation}
and factoring out the fixed gravitational wavefunction
$\chi_{\mathrm{ref}}$, the reduced universe field associated with the
four-point component takes the schematic form
\begin{equation}
    \widehat\Psi^{(4)}(z,\bar z;\hat{\mu})
    =
    \sum_J
    \int d\mu_J(\nu)\,
    \Phi_{\nu,J}^{(\hat{\mu})}(z,\bar z)\,
    a_{\nu,J},
    \label{eq:higher-dimensional-universe-field}
\end{equation}
with
\begin{equation}
    \left[
        a_{\nu,J},
        a_{\nu',J'}^\dagger
    \right]
    =
    \delta_{JJ'}\,
    \delta_{J}(\nu,\nu'),
    \qquad
    \left[
        a_{\nu,J},
        a_{\nu',J'}
    \right]
    =
    \left[
        a_{\nu,J}^\dagger,
        a_{\nu',J'}^\dagger
    \right]
    =
    0.
    \label{eq:higher-dimensional-oscillator-algebra}
\end{equation}
Here $\delta_{J}$ is the delta function with respect to the
principal-series measure $d\mu_J(\nu)$. As in the torus example, this
mode expansion determines the reduced one-universe field and its free
radial evolution without requiring each basis function to define a
complete CFT. The closure ansatz in
Eq.~\eqref{eq:deformed-CPW-map} is the minimal condition under which
the four-point harmonic data can support this reconstruction. 
If this closure fails, the bulk wavefunction space must be enlarged beyond the single-component ansatz in~Eq.\eqref{freeze_g}. 

Here the number of operator insertions should not be confused with
universe number. Correlators with $n\leq3$ insertions may still define
nontrivial one-universe wavefunctions through their dependence on the
physical gravitational variables $q$, even though conformal symmetry
fixes their insertion-point dependence on a conformally flat
boundary. The four-point sector is distinguished only because its
cross-ratio dependence supplies an explicit orthogonal harmonic basis.

The topology-changing interaction vertices contain further dynamical
information that is not fixed by the one-universe mode expansion.
They must instead be constrained by multi-boundary amplitudes,
gravitational sewing conditions, or an appropriate extension of the
bootstrap.

\section{Discussion and Outlook}\label{sec:discussion}

We have developed a third-quantized extension of AdS/CFT in which the
normalizable physical wavefunctions of connected universes form the
one-particle space of a universe Fock space, while splitting and
joining vertices generate changes of bulk connectivity. The free
evolution is inherited from reduced phase-space quantization and is
absorbed into the interaction-picture mode functions, whereas
topology-changing amplitudes provide dynamical input for the
interaction Hamiltonian.

The AdS$_3$ torus sector gives an explicit realization of this
construction. Its normalizable one-universe Hilbert space admits a
modular spectral decomposition, its radial evolution is known, and
the Cotler--Jensen wormhole supplies the leading connected
two-universe coefficient. 
We showed that its diagonal structure is compatible with a
sewing-consistent splitting interaction, thereby embedding a known
spacetime-wormhole amplitude into third-quantized dynamics. However,
the CJ amplitude alone does not determine the complete interaction
Hamiltonian; this requires additional multi-boundary gravitational
data.

\paragraph{CFT realizability and gravitational initial conditions.}

The present construction raises a broader question about the role of
AdS/CFT in selecting gravitational initial conditions. An exact CFT
supplies boundary data that determine a complete one-universe
wavefunction, but the CFT-realizable locus
$\mathfrak M_{\mathrm{CFT}}$ need not exhaust the gravitational space
$\mathfrak A_{\mathrm U}$. Thus a deformation
\begin{equation}
    \Psi_{\mathrm{CFT}}
    \longrightarrow
    \Psi_{\mathrm{CFT}}
    +
    \epsilon\,\delta\Psi,
    \qquad
    \delta\Psi\in\mathcal H_{\mathrm U},
\end{equation}
remains gravitationally admissible but need not correspond to another
consistent CFT. From the gravitational viewpoint, exact
CFT-realizability may therefore appear nongeneric, selecting a
fine-tuned locus within the larger space of admissible wavefunctions.
This differs from perturbing a state within the Hilbert space of a
fixed CFT: varying the complete one-universe wavefunction changes the
boundary-theory data themselves. Tangent directions along conformal
manifolds may remain CFT-realizable, whereas generic transverse
directions need not.\footnote{If exact boundary-CFT
consistency is fundamental, AdS/CFT may be viewed as a theory of
allowed bulk initial conditions, although it does not by itself explain
their dynamical preparation.}

Large-$N$ averaging or filtering weakens this restriction.
The smooth boundary data may select only the non-normalizable
background, while the discarded oscillatory information is carried by
the quantized normalizable sector. 
At the one-boundary level, normalizable completions with the same
filtered boundary image may then be holographically admissible without
being exact CFT realizations,
\begin{equation}
    \mathfrak M_{\mathrm{CFT}}
    \subseteq
    \mathfrak M_{\mathbb F}
    \subseteq
    \mathfrak A_{\mathrm U}.
\end{equation}
Full filtered holographic realizability additionally requires
compatibility with the surviving multi-boundary hierarchy. The size
of the gap between $\mathfrak M_{\mathrm{CFT}}$ and
$\mathfrak M_{\mathbb F}$ is discussed further in
Sec.~\ref{sec:CFT} in light of the precision-limit
estimate of Perlmutter~\cite{Perlmutter:BlackHoles}.
In this interpretation, holography constrains the background but need
not uniquely determine its normalizable completion
\cite{Kudler-Flam:2025cki,Kudler-Flam:2026nzz,Liu:2025ikq,Liu:2026fnd}.

A related tension concerns superposition. The affine gravitational
space contains
\begin{equation}
    \Psi(c_1,c_2)
    =
    \Psi_{\mathrm{bg}}
    +
    c_1\psi_1
    +
    c_2\psi_2,
    \qquad
    \psi_1,\psi_2\in\mathcal H_{\mathrm U},
    \label{eq:gravitational-wavefunction-superposition}
\end{equation}
even when a generic choice of $(c_1,c_2)$ has no exact CFT
interpretation. Positivity and integrality of the spectrum, crossing
symmetry, and higher-genus sewing are not generally preserved under
linear combinations of complete CFT data. This does not challenge
superposition within a fixed CFT; it shows instead that
$\mathfrak M_{\mathrm{CFT}}$ need not inherit the linear structure of
the gravitational Hilbert space. Third quantization accommodates the
larger gravitational space without requiring every vector to define a
conventional CFT.

The heavy--light separation of the torus example makes this distinction
concrete. Normalizable deformations leave the vacuum and light spectrum
fixed while modifying the spectrum above the black-hole threshold.
This is compatible with the proposal of Schlenker and Witten that
ensemble-like freedom may be confined to the black-hole sector, without
averaging the spectrum below threshold \cite{Schlenker:2022dyo}.
The Maloney--Witten--Keller (MWK) partition function is particularly
suggestive in this regard
\cite{Maloney:2007ud,Keller:2014xba}. It may be regarded as one
possible asymptotic boundary condition for the reduced phase-space
Schr\"odinger equation and hence as selecting a corresponding bulk
wavefunction. Although this boundary condition is obtained from a
modular sum over gravitational saddles, its continuous and partly
negative above-threshold spectral density prevents it from being the
partition function of a conventional compact unitary CFT.
It may therefore represent gravitationally admissible data outside
$\mathfrak M_{\mathrm{CFT}}$, or instead signal that the semiclassical
saddle sum is incomplete.

Exact CFT realizability may consequently be a stronger condition than
gravitational or filtered holographic realizability. Whether the
stronger condition is selected by the gravitational path integral, a
superselection rule, or third-quantized sewing and dynamics remains
open. More ambitiously, one may ask whether these principles can
determine not merely the evolution of a prescribed wavefunction, but a
preferred initial wavefunction -- or state in the universe Fock
space -- from first principles.


\paragraph{Baby universes and conditional parent amplitudes.}

The third-quantized splitting interaction provides a natural setting
in which to discuss baby-universe production. Although this
interpretation is not required for the spacetime-wormhole construction
developed above, it illustrates how a baby universe may arise as an
independent factor in the universe Fock space.

This factor should be distinguished from descriptions in which a
semiclassical Hilbert space is assigned to a specified closed universe
within a fixed holographic theory~\cite{Antonini:2023hdh}.
In the present construction, each one-universe factor is a Hilbert
space of complete gravitational wavefunctions. A baby universe
produced by a splitting interaction is therefore an additional
third-quantized universe factor, rather than a subsystem of the
microscopic Hilbert space of a fixed CFT. The question of whether its
degrees of freedom should be specified or traced over arises only
after a particular one-universe state has been selected.

We regard the incoming universe and one of the two outgoing universes
as the initial and final parent universes, respectively, and the
remaining outgoing component as the baby universe. The splitting
vertex then defines a map
\begin{equation}
    \mathcal S:
    \mathcal H_{\mathrm P,in}
    \longrightarrow
    \mathcal H_{\mathrm P,out}
    \otimes
    \mathcal H_{\mathrm{BU}}.
\end{equation}
Projecting the baby-universe leg onto a specified state
$\vert\chi\rangle_{\mathrm{BU}}$ gives the conditional parent
transition operator
\begin{equation}
    \mathcal M_\chi
    =
    {}_{\mathrm{BU}}\langle\chi\vert\mathcal S:
    \mathcal H_{\mathrm P,in}
    \longrightarrow
    \mathcal H_{\mathrm P,out}.
\end{equation}
By the canonical operator--state map, or vectorization, this operator
may equivalently be represented as
\begin{equation}
    \vert\mathcal M_\chi\rangle\!\rangle
    =
    \sum_{A,B}
    (\mathcal M_\chi)_{BA}\,
    \vert B\rangle_{\mathrm P,out}
    \otimes
    \vert A\rangle_{\mathrm P,in}^{*}.
\end{equation}
After identifying the dual incoming Hilbert space with a reflected
parent copy, this becomes a state of two parent universes. It is
entangled whenever the matrix $\mathcal M_\chi$ has rank greater than
one.

A distinguished choice is to project the baby-universe leg onto a
CFT-realizable coherent state,
\begin{equation}
    \vert\chi\rangle_{\mathrm{BU}}
    =
    \vert\Omega_{\alpha_{\mathrm{BU}}}\rangle,
\end{equation}
whose universe-field expectation value, together with the fixed
non-normalizable contribution, has boundary data corresponding to a
consistent CFT. If the incoming and outgoing parent wavefunctions are
also selected by CFT-realizable coherent states, one obtains the
conditional amplitude
\begin{equation}
    \mathcal A
    \bigl(
        \alpha_{\mathrm{out}},
        \alpha_{\mathrm{BU}};
        \alpha_{\mathrm{in}}
    \bigr)
    =
    \langle\Omega_{\alpha_{\mathrm{out}}}|
    \mathcal M_{\alpha_{\mathrm{BU}}}
    |\Omega_{\alpha_{\mathrm{in}}}\rangle,
\end{equation}
where
$\mathcal M_{\alpha_{\mathrm{BU}}}
:=
\mathcal M_{\chi=\Omega_{\alpha_{\mathrm{BU}}}}$.
This gives a candidate boundary representation of the conditional
baby-universe process in which all three universe legs are assigned
consistent boundary data.

This interpretation requires the splitting vertex to satisfy the
sewing conditions of the third-quantized theory, which ensure that an
amplitude is independent of its decomposition into elementary
splitting and joining processes. These conditions are distinct from
the ordinary CFT crossing conditions obeyed by the boundary data on
each individual universe leg. Moreover,
$\vert\mathcal M_{\alpha_{\mathrm{BU}}}\rangle\!\rangle$ is not an
ordinary state in the tensor product of two fixed-CFT Hilbert spaces:
its factors are one-universe gravitational Hilbert spaces, within
which the coherent-state parameters select CFT-realizable
wavefunctions.

Taken together, these considerations suggest that changes in universe
number and connectivity should be treated not merely as corrections to
a fixed-boundary AdS/CFT amplitude, but as part of a larger quantum
dynamics on the space of universe wavefunctions. Much of the recent
discussion of holography has focused on how spacetime emerges from a
fixed boundary quantum theory. The present construction approaches the
relation from the opposite direction: it begins with the gravitational
space of one-universe wavefunctions and asks when these wavefunctions
and their topology-changing dynamics admit a consistent
boundary-theory realization. Conventional AdS/CFT then describes a
distinguished holographic sector of the broader gravitational state
space rather than being assumed from the outset to furnish its complete
definition.

One possible boundary-theory realization of this broader description
is suggested by large-$N$ averaging or
filtering~\cite{Kudler-Flam:2025cki,Kudler-Flam:2026nzz,Liu:2025ikq,Liu:2026fnd},
supported by the no-go and precision-limit results of Gesteau and
Perlmutter discussed in the
Introduction~\cite{Gesteau:ClosedUniverses,Perlmutter:BlackHoles}.
In the torus sector, this possibility can be formulated concretely by associating the
filtered smooth data with the fixed non-normalizable background and
the oscillatory large-$N$ data with the quantized normalizable sector.
If this identification is correct, centering the normalizable sector
about its fully evolved mean leaves the background unchanged and
provides a third-quantized representation of the filter-null
one-boundary condition, while its higher
connected correlations may reproduce the filtered products of the
oscillatory data, including handle and higher-topology contributions.
The surviving hierarchy could then constrain the topology-changing
interactions through gravitational sewing and the third-quantized
Schwinger--Dyson equations.

Section~\ref{sec:higherD} sketches a possible extension to higher
dimensions by treating CFT correlation functions as boundary values
of one-universe wavefunctions and using their conformal partial-wave
decomposition as input for bulk reconstruction. A $T^2$-type
deformation may then determine the radial continuation and free
evolution of these modes. If an analogous separation between smooth
background data and quantized normalizable fluctuations can be
established, their centered correlations could likewise provide
boundary data for higher-dimensional topology-changing dynamics.
When the boundary data do not determine the complete wavefunctions or
their interactions, additional gravitational input must be retained.

The central open problem is whether these boundary and gravitational
data suffice to reconstruct a universal third-quantized Hamiltonian.
The known two-torus wormhole amplitude supplies the first
two-universe input, while higher filtered or gravitational
multi-boundary amplitudes may determine successive levels of the
Schwinger--Dyson hierarchy. Their compatibility with gravitational
sewing would provide a systematic bootstrap of the
topology-changing dynamics. A successful completion would clarify
whether ensemble-like correlations, baby-universe sectors, and the
selection of CFT-realizable initial data can emerge from a single
quantum dynamics, yielding a broader holographic description in which
the number, connectivity, and complete quantum states of universes
are themselves dynamical.

\appendix

\section{Reduced Phase-Space Quantization and York-Time Evolution}
\label{sec:CQ_York}

This appendix reviews the reduced phase-space quantization of gravity
in constant-mean-curvature (CMC) gauge and the resulting evolution in York
time~\cite{York1973}. The purpose is to explain the canonical construction underlying
the one-universe Schr\"odinger equation used in the main text. In this
formulation, the gravitational constraints are solved at the classical
level, the remaining physical degrees of freedom are coordinatized on
the reduced phase space, and the spatial volume becomes the Hamiltonian
conjugate to York time. Quantization then gives a first-order evolution
equation for physical one-universe wavefunctions.

The construction is conceptually general whenever a suitable
CMC foliation and reduced phase space exist,\footnote{The conformal CMC method extends in principle to
matter-coupled gravity. In the presence of a nonvanishing matter
momentum density, the traceless gravitational momentum is decomposed
as
$\bar\Sigma^{ij}
=
\bar\Sigma_{\mathrm{TT}}^{ij}
+
(\bar L W)^{ij}$,
where $\bar L$ is the conformal Killing operator,
$(\bar L W)_{ij}
=
\bar\nabla_iW_j+\bar\nabla_jW_i
-\frac{2}{d-1}\bar g_{ij}\bar\nabla_kW^k$,
and the vector field $W^i$ appearing in this longitudinal component
is determined by the momentum constraint. The
Hamiltonian constraint then becomes the matter-coupled
Lichnerowicz--York equation. The explicit solution of the constraints
and construction of the reduced phase space remain model dependent.} 
but its explicit implementation is highly model dependent. We first
summarize the general canonical reduction \cite{Fischer:1996qg} and then specialize to
AdS$_3$ gravity with torus spatial topology, for which the reduced
phase space, its physical Hilbert space, and the York-time evolution
can be described explicitly
\cite{Moncrief:1989dx,Hosoya:1989yj,Fujiwara:1989xg,
Ezawa:1993ti,Carlip:2004ba,Carlip:1991ij,Carlip:1994ap,
Carlip:1992cj}.

To establish our notation, we write the spacetime metric in ADM form,
\begin{equation}
    ds^2
    =
    -N^2dt^2
    +
    g_{ij}
    \left(dx^i+N^i dt\right)
    \left(dx^j+N^j dt\right).
    \label{ADM}
\end{equation}
Here and below, $d$ denotes the spacetime dimension, so that the
Cauchy slices have dimension $d-1$. 
We impose the CMC, or York, gauge
\cite{York1973} and define the normalized York time by
\begin{equation}
    \tau(t)
    :=
    -\frac{2(d-2)}{d-1}K.
    \label{eq:normalized-York-time}
\end{equation}
Equivalently,
\begin{equation}
    K
    =
    -\frac{d-1}{2(d-2)}\tau(t).
\end{equation}
Since $K$ is constant on each Cauchy slice, $\tau$ parametrizes the
CMC foliation. In $d=3$, the normalization factor equals unity and
$\tau=-K$.

The momentum constraint is
\begin{equation}
    \nabla_i
    \left(
        \frac{\pi^{ij}}{\sqrt g}
    \right)
    =
    0,
    \qquad
    \pi^{ij}
    =
    \sqrt g
    \left(
        K^{ij}-g^{ij}K
    \right).
    \label{momentum_constraints}
\end{equation}
Decomposing the extrinsic curvature into its trace and traceless parts,
\begin{equation}
    K^{ij}
    =
    \Sigma^{ij}
    +
    \frac{1}{d-1}g^{ij}K,
    \qquad
    g_{ij}\Sigma^{ij}=0,
\end{equation}
the momentum constraint reduces in CMC gauge to
\begin{equation}
    \nabla_i\Sigma^{ij}
    =
    0.
    \label{MC_York}
\end{equation}
Thus, $\Sigma^{ij}$ is transverse as well as traceless.

The reduced phase-space construction begins with the conformal
decomposition
\begin{equation}
    g_{ij}
    =
    e^{2\phi}\bar g_{ij},
\end{equation}
performed within a fixed spatial topology. Introducing the rescaled
traceless momentum
\begin{equation}
    \bar\Sigma^{ij}
    =
    e^{(d+1)\phi}\Sigma^{ij},
\end{equation}
one finds
\begin{equation}
    \bar\Sigma_{ij}
    =
    \bar g_{ik}\bar g_{jl}\bar\Sigma^{kl}
    =
    e^{(d-3)\phi}\Sigma_{ij}.
\end{equation}
The momentum constraint then becomes
\begin{equation}
    \bar\nabla_i\bar\Sigma^{ij}
    =
    0,
    \label{quadratic_diff}
\end{equation}
where $\bar\nabla_i$ is the covariant derivative associated with
$\bar g_{ij}$. The momentum constraint is therefore solved by choosing
$\bar\Sigma^{ij}$ to be transverse and traceless with respect to the
reference metric. In $d=3$, such tensors are identified with the real
parts of holomorphic quadratic differentials on the spatial Riemann
surface.

The Hamiltonian constraint determines the conformal factor through
the Lichnerowicz-York equation,
\begin{align}
    0
    =
    \mathcal H
    ={}&
    e^{2(1-d)\phi}
    \bar\Sigma_{ij}\bar\Sigma^{ij}
    -
    \frac{d-1}{4(d-2)}\tau^2
    \nonumber\\
    &-
    e^{-2\phi}
    \left[
        \bar R
        -
        2(d-2)\bar\nabla^2\phi
        -
        (d-2)(d-3)
        \bar\nabla^i\phi\bar\nabla_i\phi
    \right]
    +
    2\Lambda.
    \label{L_eqn}
\end{align}
Under suitable geometric conditions, this equation determines a
unique positive conformal factor $e^\phi$. The momentum constraint
fixes the admissible transverse-traceless momenta, while the
Hamiltonian constraint determines the remaining conformal degree of
freedom. The reduced phase space is consequently parametrized by the
conformal geometry of the spatial slice and its conjugate
transverse-traceless momentum
\cite{Fischer:1996qg}. For $d>3$, these data include the local
propagating graviton degrees of freedom. In $d=3$, where pure gravity
has no local graviton modes, the reduced phase space becomes the
cotangent bundle of the Teichm\"uller space of the spatial surface
\cite{Moncrief:1989dx}.

After imposing the CMC gauge and solving the constraints, the
Einstein--Hilbert action reduces, up to its overall normalization and
the required boundary terms, to
\begin{align}
    S_{\mathrm{EH}}
    &=
    \int dt
    \int_{\Sigma_{d-1}}d^{d-1}x
    \left[
        e^{2\phi}\sqrt g\,
        \Sigma^{ij}\dot{\bar g}_{ij}
        +
        \tau\,\partial_t\sqrt g
    \right]
    \nonumber\\
    &=
    \int dt
    \left(
        p^A\dot q_A
        +
        \tau\dot V
    \right),
    \label{eq:reduced-EH-intermediate}
\end{align}
where
\begin{equation}
    V
    :=
    \int_{\Sigma_{d-1}}d^{d-1}x\sqrt g
    =
    \int_{\Sigma_{d-1}}d^{d-1}x
    e^{(d-1)\phi}\sqrt{\bar g}
    \label{V}
\end{equation}
is the spatial volume. Integrating the second term by parts and using
$\tau$ as the evolution parameter gives
\begin{equation}
    S_{\mathrm{red}}
    =
    \int d\tau
    \left(
        p^A\frac{\partial q_A}{\partial\tau}
        -
        V(q,p;\tau)
    \right),
    \label{EH_ADM}
\end{equation}
up to the endpoint term $\tau V$. Thus, the spatial volume is the
reduced Hamiltonian generating evolution in normalized York time.

The variables $q_A$ are local coordinates on the space of conformal
geometries rather than globally defined linear mode coefficients.
Choosing a local section of the quotient by spatial diffeomorphisms
and Weyl transformations gives a family of representative metrics
$\bar g_{ij}(x;q)$. Near a reference metric, this family may be
expanded as
\begin{equation}
    \bar g_{ij}(x;q)
    =
    \bar g^{(0)}_{ij}(x)
    +
    q_A Y^{(A)}_{ij}(x)
    +
    O(q^2),
\end{equation}
where $Y^{(A)}_{ij}$ are transverse-traceless tangent vectors at
$\bar g^{(0)}$. Beyond linear order, the TT basis generally depends on
$q$, and compensating diffeomorphisms may be required to remain within
the chosen local section. No single coordinate system need cover the
entire reduced configuration space.

Only the transverse-traceless components of these variations
contribute to the reduced symplectic form. Longitudinal variations are
removed by the momentum constraint, while the trace of the momentum is
identified with York time and the conformal factor is determined by
the Hamiltonian constraint. The conjugate momenta are
\begin{equation}
    p^A
    =
    \int_{\Sigma_{d-1}}d^{d-1}x
    \sqrt{\bar g}\,
    \bar\Sigma^{ij}
    \frac{\partial\bar g_{ij}}{\partial q_A}.
\end{equation}
In $d=3$, the variables $q_A$ reduce to the moduli of the spatial
conformal geometry and $p^A$ to their conjugate momenta.

We now specialize to torus universes in asymptotically
(A)dS$_3$ spacetime. The conformal geometry of a flat torus is
parametrized by its complex modulus
\begin{equation}
    m
    =
    m_1+im_2,
    \qquad
    m_2>0,
\end{equation}
and may be represented by the unit-area metric
\begin{equation}
    d\bar s^2
    =
    \bar g_{ij}dx^idx^j
    =
    \frac{1}{m_2}
    \left|dx+m\,dy\right|^2,
    \qquad
    0\leq x,y\leq1.
    \label{eq:unit-area-torus-metric}
\end{equation}
The reference volume is therefore $\bar V=1$. The two real variables
$m_a=(m_1,m_2)$ provide coordinates on the Teichm\"uller space of the
torus, which is the upper half-plane. The physical moduli space is
obtained after quotienting by the mapping-class group
$SL(2,\mathbb Z)$.

For the flat reference metric, $\bar R=0$. Moreover,
\begin{equation}
    \bar\Sigma^2
    :=
    \bar\Sigma^{ij}\bar\Sigma_{ij}
\end{equation}
is spatially constant. The solution of the Lichnerowicz equation is
therefore also spatially constant and is given by
\begin{equation}
    e^{2\phi}
    =
    \frac{\sqrt{2\bar\Sigma^2}}
    {\sqrt{\tau^2-4\Lambda}}.
    \label{H_solved}
\end{equation}
The reduced Hamiltonian, equal to the spatial volume, becomes
\begin{equation}
    H_{\mathrm{red}}
    =
    V
    =
    \frac{
        \sqrt{m_2^2p\bar p}
    }{
        \sqrt{\tau^2-4\Lambda}
    },
    \qquad
    p
    :=
    p^1+ip^2,
    \qquad
    \bar p
    :=
    p^1-ip^2,
    \label{eq:torus-reduced-Hamiltonian}
\end{equation}
where $p^a$ are the momenta conjugate to the real moduli $m_a$.

Upon quantization, the canonical variables obey
\begin{equation}
    [m_a,p^b]
    =
    i\hbar\delta_a{}^b,
\end{equation}
and the physical wavefunction satisfies, up to operator-ordering
ambiguities,
\begin{equation}
    i\hbar
    \frac{\partial}{\partial\tau}
    \Psi(\tau,m)
    =
    \frac{1}{\sqrt{\tau^2-4\Lambda}}
    \sqrt{m_2^2p\bar p}\,
    \Psi(\tau,m).
    \label{Schrodinger}
\end{equation}
The operator $m_2^2p\bar p$ is the quantum counterpart of the
quadratic momentum on the upper half-plane. With a
mapping-class-invariant ordering, it is closely related to the
hyperbolic Laplacian on the torus moduli space.

The Euclidean continuation is implemented by rotating the lapse,
\begin{equation}
    N
    \longrightarrow
    iN,
\end{equation}
while keeping the radial foliation fixed. Since the extrinsic
curvature is inversely proportional to the lapse, this gives
\begin{equation}
    K_{ij}
    \longrightarrow
    -iK_{ij}.
\end{equation}
The Hamiltonian constraint correspondingly becomes
\begin{equation}
    0
    =
    \mathcal H_E
    =
    -K_{ij}K^{ij}
    +
    K^2
    -
    R
    +
    2\Lambda,
\end{equation}
so that the kinetic combination
$K_{ij}K^{ij}-K^2$ changes sign.

Let $\tau_E$ denote the real Euclidean York time. It is related to
the Lorentzian York time by
\begin{equation}
    \tau
    =
    i\tau_E.
    \label{eq:Lorentzian-Euclidean-York-time}
\end{equation}
The Euclidean Lichnerowicz equation then gives
\begin{equation}
    e^{2\phi}
    =
    \frac{\sqrt{2\bar\Sigma_E^2}}
    {\sqrt{\tau_E^2+4\Lambda}}.
    \label{H_solved_E}
\end{equation}
After continuing both the reduced Hamiltonian and York time, their
factors of $i$ cancel in the evolution equation. The Euclidean
wavefunction therefore satisfies
\begin{equation}
    i\hbar
    \frac{\partial}{\partial\tau_E}
    \Psi_E(\tau_E,m)
    =
    \frac{1}{\sqrt{\tau_E^2+4\Lambda}}
    \sqrt{m_2^2p\bar p}\,
    \Psi_E(\tau_E,m).
    \label{E_Schrodinger}
\end{equation}
Thus, in this reduced formulation, the Euclidean radial evolution
retains the same first-order Schr\"odinger form as the Lorentzian
evolution.

The explicit York-time dependence of the reduced Hamiltonian can be
absorbed into a redefinition of the evolution parameter. For
Lorentzian AdS$_3$, with $\Lambda=-1/\ell^2$, the corresponding
Wheeler--DeWitt (WdW) time is
\begin{equation}
    T(\tau)
    :=
    \int_0^\tau
    \frac{d\tau'}{\sqrt{\tau'^2+4/\ell^2}}
    =
    \operatorname{sinh}^{-1}
    \left(
        \frac{\ell\tau}{2}
    \right),
    \label{eq:WdW-time-definition}
\end{equation}
where the integration constant has been chosen so that $T(0)=0$.
Thus, the dependence on the cosmological constant and York time is
encoded entirely in the relation $ \tau=\frac{2}{\ell}\sinh T$.

For Euclidean AdS$_3$, the corresponding WdW time is
\begin{equation}
    T_E(\tau_E)
    :=
    \int_{2/\ell}^{\tau_E}
    \frac{d\tau_E'}{\sqrt{\tau_E'^2-4/\ell^2}}
    =
    \operatorname{cosh}^{-1}
    \left(
        \frac{\ell\tau_E}{2}
    \right).
    \label{eq:Euclidean-WdW-time}
\end{equation}
The lower limit is fixed geometrically: the CMC slice at
$\tau_E=2/\ell$ approaches the asymptotic AdS boundary, where
$T_E=0$. As $\tau_E$ increases, the foliation evolves radially toward
the Euclidean Rindler horizon, reached in the limit
$\tau_E\to\infty$, or equivalently $T_E\to\infty$. The Euclidean WdW
time therefore ranges over $0\leq T_E<\infty$ and parametrizes radial
evolution from the AdS boundary to the Rindler horizon.

Using the chain rule, Eq.~\eqref{E_Schrodinger} becomes
\begin{equation}
    i\hbar
    \frac{\partial}{\partial T_E}
    \Psi_E(T_E,m)
    =
    \widehat h_{\mathrm{mod}}
    \Psi_E(T_E,m),
    \qquad
    \widehat h_{\mathrm{mod}}
    :=
    \sqrt{m_2^2p\bar p}.
    \label{eq:Euclidean-WdW-time-Schrodinger}
\end{equation}
Thus, the cosmological-constant dependence is absorbed into the
relation between $T_E$ and $\tau_E$, while
$\widehat h_{\mathrm{mod}}$ acts only on the torus moduli. Since the
remainder of this paper concerns Euclidean radial evolution, we
henceforth write $T\equiv T_E$ and suppress the subscript $E$ on the
wavefunction.



\section*{Acknowledgments}

SH would like to thank Animik Ghosh and Masaki Shigemori for discussions and the department of mathematics at Nagoya University for their hospitality during his visits where part of this work was done.


\end{document}